\documentclass{emulateapj}
\usepackage{natbib}
\usepackage{epsfig}

\newcommand{\Msun}{\mathrm{M}_{\odot}}
\newcommand{\Lsun}{\mathrm{L}_{\odot}}

\def\Mpc2{\; {\rm M}_{\odot}\, {\rm pc}^{-2}}
\newcommand{\fsfr}{f_{\rm sfr}}
\newcommand{\age}{\tau}   

\begin{document}

\slugcomment{Submitted to ApJ}
\shortauthors{Orban et al.}
\shorttitle{Star Formation Histories of Dwarf Galaxies}

\title{Delving Deeper into the Tumultuous Lives of Galactic Dwarfs: \\ 
       Modeling Star Formation Histories}
        
\author{ Chris Orban\altaffilmark{1}$^,$\altaffilmark{2}, 
         Oleg Y. Gnedin\altaffilmark{3}, 
         Daniel R. Weisz\altaffilmark{4}, 
         Evan D. Skillman\altaffilmark{4}, \\
         Andrew E. Dolphin\altaffilmark{5}, and 
         Jon A. Holtzman\altaffilmark{6}}

\altaffiltext{1}{Center for Cosmology and Astro-Particle Physics,
    The Ohio State University, 191 W Woodruff Ave, Columbus, OH 43210}
\altaffiltext{2}{Department of Physics,
    The Ohio State University, 191 W Woodruff Ave, Columbus, OH 43210; \mbox{\tt  orban@mps.ohio-state.edu} }
\altaffiltext{3}{Department of Astronomy, University of Michigan, 
    500 Church St., Ann Arbor, MI 48109; \mbox{\tt ognedin@umich.edu} }
\altaffiltext{4}{Department of Astronomy, 
    University of Minnesota, 116 Church St. SE, Minneapolis, MN 55455}
\altaffiltext{5}{Raytheon Corporation, USA}
\altaffiltext{6}{Astronomy Department
                 New Mexico State University, Box 30001, MSC 4500
                 Las Cruces, NM 88003}

\date{\today}

\begin{abstract}
The paucity of observed dwarf galaxies in the Local Group relative to
the abundance of predicted dark matter halos remains one of the
greatest puzzles of the $\Lambda$CDM paradigm.  Solving this puzzle
now requires not only matching the numbers of objects but also
understanding the details of their star formation histories.  We
present a summary of such histories derived from the HST data using
the color-magnitude diagram fitting method.  To reduce observational
uncertainties, we condense the data into five cumulative parameters --
the fractions of stellar mass formed in the last 1, 2, 5, and 10 Gyr,
and the mean stellar age.  We interpret the new data with a
phenomenological model based on the mass assembly histories of dark
matter halos and the Schmidt law of star formation.  The model
correctly predicts the radial distribution of the dwarfs and the
fractions of stars formed in the last 5 and 10 Gyr.  However, in order
to be consistent with the observations, the model requires a
significant amount of recent star formation in the last 2 Gyr.  Within
the framework of our model, this prolonged star formation can be
achieved by adding a stochastic variation in the density threshold of
the star formation law.  The model results are not sensitive to late
gas accretion, the slope of the Schmidt law, or the details of cosmic
reionization.  A few discrepancies still remain: our model typically
predicts too large stellar masses, only a modest population of
ultra-faint dwarfs, and a small number of dwarfs with anomalously
young stellar populations.  Nevertheless, the observed star formation
histories of Local Group dwarfs are generally consistent the expected
star formation in cold dark matter halos.
\end{abstract}

\keywords{cosmology: theory --- galaxies: formation --- galaxies: dwarf}

\section{Introduction}
  \label{sec:intro}

The discrepancy between the number of observed dwarf satellite
galaxies of the Milky Way and the anticipated number of dark matter
halos in cosmological simulations has been heralded as the ``missing
satellite problem'' \citep{Klypin1999, Moore1999, kauffmann_etal93}
and it still remains one of the greatest puzzles of the $\Lambda$CDM
paradigm.  One class of suggested solutions to this puzzle involves
modifications to the nature of dark matter
\citep[e.g.,][]{spergel_steinhardt01} or to the initial conditions of
cosmic structure \citep[e.g.,][]{ZentnerBullock2003a,
ZentnerBullock2003b, KamionkowskiLiddle2000, Colin_et_al_2000}, while
another class of solutions invokes astrophysical arguments such as
inefficient cooling of cosmic gas and/or feedback from young stars
\citep[e.g.][]{ThoulWeinberg96, DekelWoo2003}. 

Motivated by this latter class of solutions, \citet*[][hereafter
referred to as KGK04]{tumultuous04} developed a star formation model
for the satellite galaxies, based on the mass assembly histories of
individual dark matter halos in a $\Lambda$CDM cosmological
simulation.  That model incorporated the accretion of gas in
hierarchical mergers and the loss of gas caused by the extragalactic
UV background, and included both a continuous mode and a starburst
mode of star formation.  The model correctly reproduced the observed
number of satellites of the Milky Way and M31, the radial distribution
of luminous satellites, and the morphological segregation of dwarf
irregular (dIrr) and dwarf spheroidal (dSph) galaxies.  Recent
$N$-body simulations of \citet*{Madau_Diemand_Kuhlen_2008}, with the
updated WMAP3 cosmological parameters, have confirmed the qualitative
picture of KGK04.

The amount of observational data for Local Group dwarfs has been 
increasing steadily in the
last several years, shifting the focus from simply counting the number
of dwarf galaxies to deriving their detailed star formation histories
(SFHs).  Most Local Group dwarfs have now been observed with the {\it
HST}, which has provided color-magnitude diagrams of resolved stellar
populations in single or multiple fields \citep{2005Dolphin_et_al,
Holtzman_et_al_2006}.  We parametrize these data as the fractions of
stars formed in the last 1~Gyr, 2~Gyr, 5~Gyr, and 10~Gyr, and present
these fractions in our Table \ref{tab:sfdata_lg}.  The data show a
great variety of star formation histories -- some continuous, some
bursty, some truncated.  In this paper we extend the model of KGK04 to
account for such varied histories of the Local Group dwarfs and argue
that adding a stochastic threshold to the star formation law greatly improves
the comparison between models and observations.
We also make predictions for the number of ultra-faint satellites of 
the Milky Way and M31 of which there have been many recent discoveries
with the Sloan Digital Sky Survey (SDSS) and the MegaCam survey
\citep{Belokurov2006, Belokurov2007, Irwin2007, Liu_etal08, Martin2006,
SimonGeha2007, Walsh2007, Willman2005a, Willman2005b, Zucker2004,
Zucker2006b, Zucker2006, Zucker2007}.

Our approach is complementary to running full hydrodynamic simulations
of galaxy formation.  We attempt to model star formation with a
phenomenological recipe, containing several but not too many free
parameters.  We investigate the effects of varying each of these
parameters and constrain them by comparing the model predictions 
for each parameter set with the observations.  In this way, 
we obtain empirical rules that
govern star formation in dwarf galaxies, even though we may not yet
fully understand their physical origin.  In an alternative approach,
one uses cosmological hydrodynamic simulations with a detailed
prescription for gas cooling and star formation, and thus obtains more
physically sound results. Discrepancies between the 
simulation results and the data may, however, be difficult to interpret
or computationally expensive to investigate. By combining the two approaches, 
our empirical rules will guide future developments of more sophisticated 
models of galaxy formation.

The paper is organized as follows.  We give a brief summary of the
observational data in \S\ref{sec:obs}.  We describe our model in
\S\ref{sec:model} and report the primary results from adding
stochasticity to the star formation law in \S\ref{sec:threshold}.  In
\S\ref{sec:other} we describe the results of several variants of our
star formation model, including late accretion of gas, steeper slopes
of the Schmidt law, an extended epoch of reionization, and rejection
of galaxies with delayed star formation; none of these changes
significantly alters our results.  In \S\ref{sec:lowmass} we show our
model projections for the new low-mass dwarfs, and in
\S\ref{sec:conclusions} we summarize our main results.

\begin{table*}
\begin{center}
\caption{ Star Formation Histories of Satellite Galaxies of MW and M31
   \label{tab:sfdata_lg}}
\begin{tabular}{lcccrcllllr}
\tableline\tableline\\
\multicolumn{1}{l}{Galaxy} &
\multicolumn{1}{c}{Alternate Name} &
\multicolumn{1}{c}{Type} &
\multicolumn{1}{c}{Host} &
\multicolumn{1}{c}{$r_{\rm host}$ (kpc)} &
\multicolumn{1}{c}{log$(M_{*,\sun})$} &
\multicolumn{1}{c}{$f_{1G}$} &
\multicolumn{1}{c}{$f_{2G}$} &
\multicolumn{1}{c}{$f_{5G}$} &
\multicolumn{1}{c}{$f_{10G}$} &
\multicolumn{1}{c}{$\age$ (Gyr)}
\\[2mm] \tableline\\
\noindent M33    &  NGC598        & Sc    & M31 & 203  & 9.9 & 0.093  &        &       & 0.52  & 8.4       \\
LMC     &               & Irr   & MW  & 50   & 9.7 & 0.078  & 0.17   & 0.42  & 0.70  & 6.7        \\
SMC &                   & Irr   & MW  & 63   & 9.2 & 0.096  & 0.18   & 0.48  & 0.65  & 6.6        \\  
M32 &  NGC221           & dE    & M31 & 6    & 9.1 & 0.042  &        &       & 0.50  & 8.5       \\
NGC205 & M110           & dE    & M31 & 58   & 9.0 & 0.0049 & 0.0050 & 0.055 & 0.48  & 10.5      \\ 
IC10 &                  & dIrr  & M31 & 255  & 8.7 & 0.060  & 0.14   & 0.52  & 0.75  & 7.1       \\ 
NGC6822 &               & dIrr  & MW  & 500  & 8.7 & 0.087  & 0.16   & 0.57  & 0.68  & 6.9       \\ 
NGC3109 &               & dIrr  & MW  & 1360 & 8.7 & 0.054  & 0.065  & 0.084 & 0.12  & 11.1      \\  
NGC185 &                & dSph  & M31 & 175  & 8.6 & 0.0053 & 0.0053 & 0.090 & 0.51  & 10.5       \\ 
NGC147 &                & dSph  & M31 & 101  & 8.4 & 0.032  & 0.036  & 0.050 & 0.17  & 12.4        \\  
IC1613 &                & dIrr  & M31 & 505  & 8.3 & 0.059  & 0.11   & 0.42  & 0.64  & 7.7       \\ 
WLM  &  DDO221          & dIrr  & M31 & 840  & 8.2 & 0.14   & 0.35   & 0.55  & 0.69  & 6.7       \\  
Sex B & DDO70           & dIrr  & MW  & 1320 & 8.2 & 0.049  & 0.067  & 0.11  & 0.21  & 11.1    \\ 
Sex A & DDO75           & dIrr  & MW  & 1440 & 7.9 & 0.15   & 0.29   & 0.38  & 0.41  & 9.3      \\ 
Sagittarius &           & dSph  & MW  & 28   & 7.7 & 0.0008 & 0.0008 & 0.52  & 0.86  & 6.5     \\ 
Fornax &                & dSph  & MW  & 138  & 7.5 & 0.013  & 0.059  & 0.33  & 0.73  & 7.4     \\ 
UGC4879 & VV124         & dIrr  & MW  & 1100 & 7.3 \\
Pegasus & DDO216        & dIrr  & M31 & 410  & 7.2 & 0.057  & 0.095  & 0.40  & 0.64  & 7.4      \\ 
UGCA92 & EGB\_0427+63   & dIrr  & MW  & 1300 & 7.2 &        &        &       &       &          \\ 
Sag DIG & ESO594-4      & dIrr  & MW  & 1060 & 7.1 & 0.11   & 0.17   & 0.20  & 0.20  & 11.6    \\ 
AndVII & Cassiopia dSph & dSph  & M31 & 216  & 7.1 & 0.016  & 0.022  & 0.022 & 0.025 & 12.9      \\
AndI &                  & dSph  & M31 & 48   & 7.1 & 0.0038 & 0.0098 & 0.087 & 0.67  & 8.9    \\ 
AndII  &                & dSph  & M31 & 160  & 7.0 & 0.0049 & 0.0086 & 0.076 & 0.50  & 9.2      \\ 
AndVI  & Pegasus dSph   & dSph  & M31 & 266  & 6.9 & 0.0023 & 0.023  & 0.19  & 0.60  & 9.0      \\ 
Leo A  & DDO69          & dIrr  & MW  & 800  & 6.8 & 0.13   & 0.31   & 0.65  & 0.78  & 6.2     \\ 
Antlia  &               & dSph  & MW  & 1330 & 6.8 & 0.043  &        &       & 0.43  & 9.0    \\
LeoI  & DDO74           & dSph  & MW  & 270  & 6.8 & 0.0099 & 0.18   & 0.50  & 0.76  & 6.4      \\ 
Aquarius & DDO210       & dIrr  & MW  & 950  & 6.7 & 0.037  & 0.083  & 0.12  & 0.12  & 12.0    \\ 
AndIII  &               & dSph  & M31 & 68   & 6.5 & 0.0022 & 0.0061 & 0.10  & 0.47  & 9.8     \\ 
Cetus  &                & dSph  & M31 & 680  & 6.4 & 0.0045 & 0.013  & 0.17  & 0.52  & 9.9     \\ 
LGS3  & Pisces          & dIrr  & M31 & 284  & 6.3 & 0.015  & 0.046  & 0.16  & 0.43  & 9.8       \\ 
LeoII & DDO93           & dSph  & MW  & 205  & 6.3 & 0.0028 & 0.012  & 0.025 & 0.70  & 8.8     \\ 
Phoenix  &              & dIrr  & MW  & 405  & 6.3 & 0.027  & 0.071  & 0.23  & 0.42  & 10.3     \\ 
Sculptor  &             & dSph  & MW  & 88   & 6.3 & 0.010  & 0.016  & 0.026 & 0.14  & 12.6      \\ 
Tucana &                & dSph  & MW  & 870  & 6.2 & 0.0048 & 0.011  & 0.014 & 0.30  & 11.6     \\ 
AndXV   &               & dSph  & M31 & 170  & 6.2 \\
AndXVI  &               & dSph  & M31 & 270  & 6.1 \\
Sextans &               & dSph  & MW  & 86   & 6.1 & 0.00   & 0.00   & 0.00  & 0.00  & 12.0     \\
AndV &                  & dSph  & M31 & 117  & 6.0 & 0.0045 & 0.048  & 0.066 & 0.35  & 10.8      \\ 
Carina &                & dSph  & MW  & 94   & 6.0 & 0.0065 & 0.0077 & 0.43  & 0.67  & 7.1      \\ 
Draco &  DDO208         & dSph  & MW  & 79   & 5.9 & 0.0004 & 0.010  & 0.025 & 0.49  & 10.9     \\ 
Ursa Minor & DDO199     & dSph  & MW  & 69   & 5.9 & 0.00   & 0.00   & 0.00  & 0.00  & 12.0    \\
AndX &                  & dSph  & M31 & 110  & 5.9 &        &        &       &       &         \\ 
AndXIV  &               & dSph  & M31 & 162  & 5.8 \\
AndXVII   &             & dSph  & M31 & 44   & 5.8 &        &        &       &       &         \\ 
AndIX   &               & dSph  & M31 & 45   & 5.7 &        &        &       &       &         \\ 
\tableline              
\end{tabular}
\end{center}
{\small {\sc Notes.}---Listed are all presently known satellite
galaxies of the MW and M31 within 1 $h^{-1}$ Mpc of either host, with
stellar mass $M_* > 5 \times 10^5\, \Msun$.  Star formation data are
shown if available -- here we parameterize the SFH in terms of the
fraction of total stellar mass formed in the last 1, 2, 5, and 10 Gyr
(i.e., $f_{1G}$, $f_{2G}$, $f_{5G}$, $f_{10G}$) and the mean
mass-weighted stellar age, $\age$.  Most of the star formation data
are derived from the Local Group Stellar Photometry Archive
\citep{Holtzman_et_al_2006}.  Exceptions are M32, M33, Antlia, Sextans
and Ursa Minor, which are taken from \citet{2005Dolphin_et_al} and
$\age$ is estimated from the reported $f_{10G}$ value.  The LMC data
are inferred from \citet{SmeckerHane2002}; SMC from
\citet{HarrisZaritsky2004}.  Distances from the host, $r_{\rm host}$,
are directly taken or inferred from \citet{Grebel_et_al_2003}
 with some exceptions:  AndX \citep{Zucker2007},
AndXIV \citep{majewski_etal07}, AndXV and AndXVI \citep{ibata_etal07},
AndXVII \citep{Irwin_et_al_2008}, UGC4879 \citep{kopylov_etal08}, UGCA 92 \citep{Mateo1998},
and M33 \citep{McConnachie_etal2004}.
Stellar mass, $M_*$, is from \citet{DekelWoo2003}, except for M32 and
Sagittarius, whose values are estimated using $M_*/L_V = 3\
\Msun/\Lsun$, with $L_V$ from \citet{Mateo1998}; similarly we estimate
$M_*$ for AndIX, AndX, AndXIV, AndXV, AndXVI, AndXVII and UGC 4879 from
the quoted $L_V$ or $M_V$ in the same references cited for $r_{\rm
host}$.
  \vspace{0.5cm}}
\end{table*}

\section{Observed Star Formation Histories}
  \label{sec:obs}

The star formation data used in this study, and listed in
Table~\ref{tab:sfdata_lg}, come primarily from Hubble Space Telescope
observations with the WFPC2 camera.  For a majority of the dwarfs, the
SFHs were measured by the color-magnitude diagram fitting method,
described in \citet{Dolphin2002}, using photometric data from the
Local Group Stellar Photometry
Archive\footnote[1]{http://astronomy.nmsu.edu/holtz/archival/html/lg.html}
\citep{Holtzman_et_al_2006}.  We condense the data into five
parameters -- the fractions of stellar mass formed in the last 1, 2,
5, and 10 Gyr, respectively ($f_{1G}$, $f_{2G}$, $f_{5G}$, $f_{10G}$),
and the mass-weighted mean age of the stellar population ($\age$).
The HST observations included in archive are far from homogeneous,
and, therefore, the uncertainties in the derived star formation
histories cover a broad range.  However, by concentrating on
cumulative star formation fractions the uncertainties are greatly
reduced, compared to specific star formation rates at specific times.
Thus, the entries in Table~\ref{tab:sfdata_lg} do not have associated
errors.  Constraining comparisons between models and observations to
cumulative distributions greatly reduces the sensitivity of these
comparisons to uncertainties in individual galaxies.

Note that the observed {\it HST} fields do not necessarily cover the
full extent of each galaxy and, therefore, do not allow us to calculate
the total stellar mass.  For the estimate of total stellar mass, with
a few exceptions, we take the values quoted in \citet{DekelWoo2003}.

The rest of the SFHs are taken from a variety of sources, as indicated
in the notes to Table \ref{tab:sfdata_lg}.  In the case of the LMC, the
star formation data are inferred from \citet{SmeckerHane2002}.  Since
they give the star formation rates per unit area, in units of $\Msun\,
{\rm yr}^{-1}\, {\rm deg}^{-2}$, separately for the disk and the bar
of the LMC, we assume that the extent of the disk is 3.5 times larger
than the extent of the bar in determining the star formation fractions
and the mean age (A.~A. Cole, personal communication).

The star formation histories for five dwarfs (M32, M33, Antlia,
Sextans, Ursa Minor) are taken from Table 1 of
\citet{2005Dolphin_et_al}.  Since that table does not include the
fraction of a galaxy's star formation in the last $2$ Gyr or $5$ Gyr,
our table is also missing these entries, except for Sextans and Ursa
Minor.  \citet{2005Dolphin_et_al} reported that Sextans and Ursa Minor
have formed no stars in the past 10 Gyr (i.e., $f_{10G} = 0$), which
implies that the fractions of stars formed in the last 1~Gyr, 2~Gyr,
and 5~Gyr are also consistent with zero.

For the purpose of comparing the radial distributions of different
morphological types with model predictions, we combine the Sc, Irr,
and dIrr types into one broadly-defined ``dIrr'' group, and the dE,
dSph types into the ``dSph'' group.

One object not included in Table~\ref{tab:sfdata_lg} is the Canis Major
dwarf, discovered by \citet{martin_etal04} in the SDSS field, which
could be part of a larger Monoceros tidal stream.  This object is the
closest discovered to the Galaxy, at a heliocentric distance of only 8
kpc.  Whether this galaxy is still gravitationally self-bound is
unclear at present \citep{butler_etal07}, but in any case it is
being strongly tidally disrupted.  An additional reason for not
including it in our comparison is that such an object would have been
completely disrupted in the $N$-body numerical simulation used to
construct our models.

Note also that Table~\ref{tab:sfdata_lg} does not include any of the
newly-discovered SDSS and MegaCam ultrafaint dwarfs, since they are
likely to have stellar masses below $5~\times~10^5\,~\Msun$.  We list
these new objects in Table~\ref{tab:lowmass_tbl} and compare their
estimated stellar mass function with the predictions of our fiducial
model in \S\ref{sec:lowmass}.

\begin{figure*}[p]
\vspace{-1.0cm}
\centerline{\epsfig{file=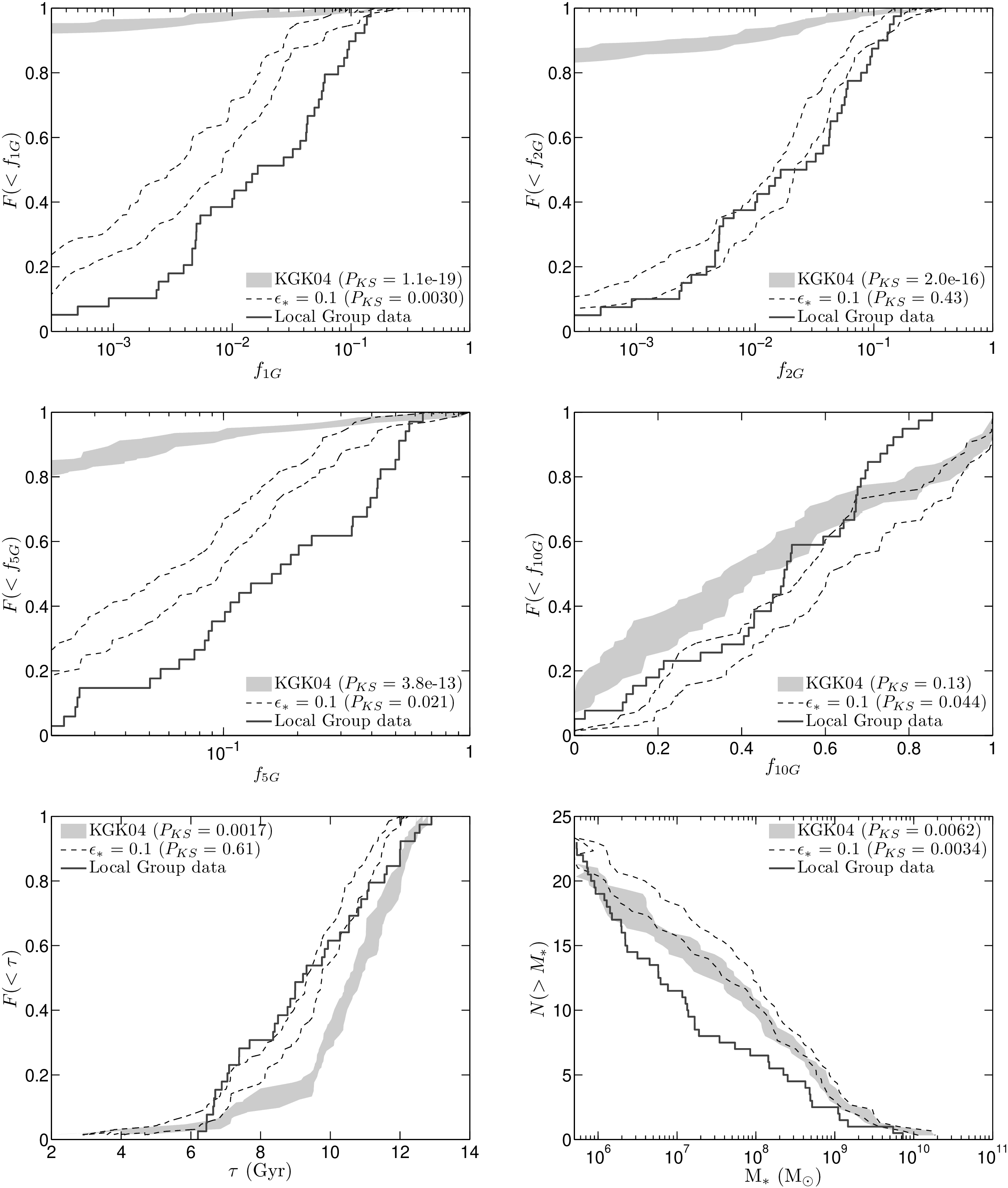, angle=0, width=7.0in}}
\vspace{-1.6cm}
\caption{{\it Top four panels:} Cumulative distributions of the
fractions of stellar mass formed in the last 1, 2, 5, and 10 Gyr.  The
$f_{1G}$ and $f_{2G}$ fractions reflect recent star formation, while
the $f_{5G}$ and $f_{10G}$ fractions represent the overall star
formation history.  Solid line shows the data for the Local Group.
Gray shaded region shows the spread of predictions of the KGK04 model,
for 10 random realizations of the model.  Dashed region shows the
range predicted by our model with a stochastic star formation
threshold, $\epsilon_* = 0.1$, also for 10 realizations.  The numbers
in parentheses show the Kolmogorov-Smirnov probability of the model
average being consistent with the data.  {\it Bottom left:} cumulative
distribution of the mass-weighted mean age of stellar population.
{\it Bottom right:} Cumulative stellar mass function per host halo.}
  \label{fig:sfplots}
\end{figure*}

\begin{figure*}[p]
\vspace{-0.2cm}
\centerline{\epsfig{file=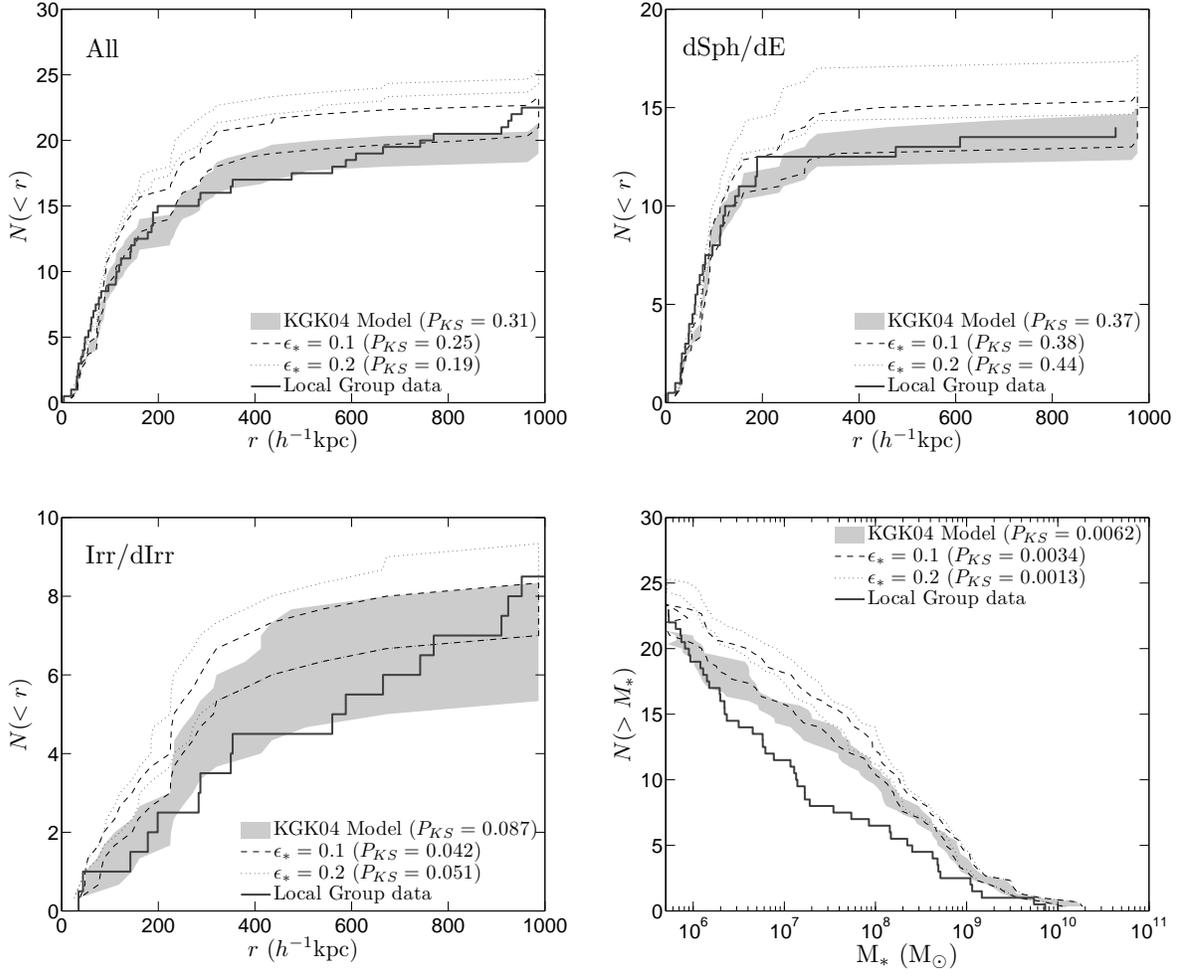, angle=-0, width=7.39in}}
\vspace{-0.8cm}
\caption{Three cumulative radial plots and a cumulative stellar mass
function comparing the data for observed Local Group dwarfs (solid
line) to three variants of the star formation model.  The KGK04 model
(gray shaded region) does not include stochasticity in the star
formation density threshold, $\Sigma_{\rm th}$, (i.e., $\epsilon_* =
0$) whereas the two other models include stochasticity with a
logarithmic dispersion $\epsilon_* = 0.1$ (contours of dotted lines)
and $\epsilon_* = 0.2$ (contours of dashed lines).  The other
parameters of the Kennicutt-Schmidt law are kept the same for the
three models.  The models include an intrinsic spread from the random
assignment of the angular momentum spin parameters to the halos.  This
spread is shown here as a filled region, or contours outlining a
region, which encompasses the range of model predictions for 10 random
realizations of the model.}
  \label{fig:radial}
\end{figure*}

\begin{table*}[p]
\begin{center}
\caption{ \sc Stochastic Threshold of Star Formation Law
   \label{tab:sfdata}}
\begin{tabular}{lccccc}
\tableline\tableline\\
\multicolumn{1}{c}{} &
\multicolumn{1}{c}{$\epsilon_* = 0$} &
\multicolumn{1}{c}{$\epsilon_* = 0.05$} &
\multicolumn{1}{c}{$\epsilon_* = 0.1$} &
\multicolumn{1}{c}{$\epsilon_* = 0.15$} &
\multicolumn{1}{c}{$\epsilon_* = 0.2$} 
\\[2mm] \tableline\\
$P_{KS}$: $r_{\rm all}$    & 3.1e-1  & 3.0e-1 & 2.5e-1  & 2.1e-1 & 2.2e-1 \\
$P_{KS}$: $r_{\rm dSph}$   & 3.7e-1  & 3.6e-1 & 3.8e-1  & 3.6e-1 & 4.9e-1 \\
$P_{KS}$: $r_{\rm dIrr}$   & 8.7e-2  & 5.2e-2 & 4.2e-2  & 3.3e-2 & 3.9e-2 \\
$P_{KS}$: $f_{1G}$         & 1.1e-19 & 2.0e-6 & 3.0e-3  & 3.1e-2 & 1.1e-1 \\
$P_{KS}$: $f_{2G}$         & 2.0e-16 & 3.4e-2 & 4.3e-1  & 3.6e-1 & 1.0e-1 \\
$P_{KS}$: $f_{5G}$         & 3.8e-13 & 2.5e-4 & 2.1e-2  & 7.8e-2 & 2.9e-1 \\
$P_{KS}$: $f_{10G}$        & 1.3e-1  & 9.9e-2 & 4.4e-2  & 1.6e-2 & 2.5e-3 \\
$P_{KS}$: $\age$           & 1.7e-3  & 1.3e-1 & 6.1e-1  & 3.8e-1 & 1.5e-1 \\
$P_{KS}$: $M_*$            & 6.2e-3  & 3.1e-3 & 3.4e-3  & 2.8e-3 & 1.2e-3 \\
Dwarfs per halo            & 20      & 21     & 22      & 24     & 26     \\
\tableline
\end{tabular}
\end{center}
{\small
\begin{center}
{\sc Notes.}---Other fixed parameters are $\Sigma_{\rm th0} = 5
\Mpc2$, $\fsfr = 1$.  The observed \\
number in the Local Group is 46/2 = 23 dwarfs per host halo.
\end{center}
}
\end{table*}


\section{Star Formation Model}
  \label{sec:model}

The KGK04 model of star formation in dwarf galaxies is
based on the mass assembly history of dark matter halos in a
collisionless $\Lambda$CDM simulation of the Local Group-like
environment.  The simulation volume contains three large host halos,
with virial masses $(1.2-1.7) \times 10^{12}\ h^{-1}\ \Msun$ at $z=0$,
resolved with $\sim 10^6$ dark matter particles.  The model
incorporates the accretion of gas in hierarchical mergers and the loss
of gas caused by the extragalactic UV background (following the
filtering mass approach of \citealt{Gnedin2000}), and includes both a
continuous mode and a starburst mode of star formation.

The model assumes that in the satellite halos the accreted gas
dissipates its energy via radiative cooling and forms a disk.  The
surface density of the gas follows an exponential profile,
\begin{equation}
  \Sigma_{g}(r) = \Sigma_0 \exp{(-r / r_d)},
  \label{eq:gas_radial}
\end{equation}
with the scale length $r_d$ set by the satellite halo's virial radius
$r_{\rm vir}$ and angular spin parameter $\lambda$:
\begin{equation}
  r_d = \lambda \, r_{\rm vir} \, 2^{-1/2} \, 
        \exp{\left[{c (V_4/V_{\rm max})^2}\right]}.
  \label{eq:r_d}
\end{equation}
The last factor accounts for the less efficient gas dissipation in
small halos with the virial temperature $T_{\rm vir} \lesssim {\rm
few} \times 10^4$ K, or equivalently, maximum circular velocity
$V_{\rm max} \lesssim 50$ km s$^{-1}$.  This factor is written in
terms of $V_4 \equiv 16.7$ km s$^{-1}$, the virial velocity
corresponding to the virial temperature $T_{\rm vir} = 10^4$ K, and a
constant, $c = 10$, chosen to reproduce the correct total number of
dwarfs.  This important factor, coupled with the density threshold for
star formation, suppresses star formation in most low-mass halos.
Equation (\ref{eq:r_d}) is consistent with results of recent
hydrodynamic simulations of galaxy formation
\citep{KravstovGnedin2005, tassis_etal08}.  An alternative
parametrization of dwarf gaseous disks with a temperature floor at
$\sim 10^4$ K is given by \citet{kaufmann_etal07}.  In their model the
disks are vertically puffed-up, which works to the same effect to
reduce the gas density.  The observed stellar core radii in the Local
Group dwarfs are a factor of $2-3$ smaller than those predicted by our
model, which may favor the vertical expansion over the radial one
assumed here.  However, since it would not change the surface density
$\Sigma_{g}$, we cannot fit it in our framework of the
Kennicutt-Schmidt law of star formation.

The spin parameter $\lambda$ is drawn randomly from a standard
log-normal distribution with $\bar{\lambda} = 0.045$ and
$\sigma_\lambda = 0.56$ \citep[][]{Vitvitska02, Hernandez07}.  This
adds an intrinsic variance in the predictions of the star formation
model.  While the KGK04 model used only one set of the
randomly-selected $\lambda$ values to compare with the observations,
in this paper we take 10 random realizations of each model, in order
to account for this intrinsic variance.

The gaseous disk is modeled spatially by 50 radial zones.  At each
simulation output epoch (about every $10^8$ yr), the newly accreted
gas is added to these radial zones, according to the exponential
profile of eq. (\ref{eq:gas_radial}).  In each zone, the rate of star
formation is determined by the Kennicutt-Schmidt law:
\begin{equation}
  \dot{\Sigma}_{*} = 2.5 \times 10^{-4} \fsfr
     \left({\Sigma_{g} \over 1 \Mpc2}\right)^n 
     \ \Msun \, {\rm kpc}^{-2} \, {\rm yr}^{-1},
\end{equation}
wherever the gas density, $\Sigma_{g}$, exceeds the threshold,
$\Sigma_{\rm th}$ \citep{kennicutt98}.  Standard parameters are $n
\approx 1.4$, $\fsfr = 1$, and $\Sigma_{\rm th} = 5 \, \Mpc2$.  Most
of the variants of the model discussed in this paper employ this star
formation law but with an important addition, described in
\S\ref{sec:threshold} below.

In addition to this quiescent mode of star formation, the model also 
allows a starburst mode prompted by strong tidal interactions with other
halos (see section 6.1 in KGK04).  In this mode there is no density
threshold and even galaxies with $\Sigma_{g} < \Sigma_{\rm th}$
can form stars if a strong enough interaction occurs.

Finally, the effect of stellar evolution is taken into account
following \citet{prieto_gnedin06}.  They find that some 40\% of the
initial stellar mass is lost to stellar wind and supernovae after
$5-10$ Gyr, given the assumptions of the initial stellar mass function
from \citet{kroupa01}, stellar remnant masses as a function of initial
stellar mass from \citet{chernoff_weinberg90}, and main sequence
lifetimes from \citet{hurley_etal00}.  In the models investigated here
we simply assume that 40\% of the initial stellar mass is lost between
the time when the stars formed and the present day.  We do not
``recycle'' the liberated gas back into the ISM, which would make it
available to form more stars. As a result our estimate of the stellar 
masses are the lower limit of the true masses -- a simple decrease
of the total stellar mass, $M_*$, for each dwarf by a factor of $0.6$.
Accordingly, the $f_{[1,2,5,10]G}$ and $\tau$ values are unchanged 
by stellar evolution in our models and since the {\it HST} photometry
for the Milky Way satellites is typically good enough to measure
the main sequence stars we simply compare the observed 
$f_{[1,2,5,10]G}$ and $\tau$ values (Table~\ref{tab:sfdata_lg}) to the
 same quantities for the simulated dwarfs without any corrections.

We mark as luminous (having corrected $M_*$ for stellar evolution as
 just described) those satellite halos with a predicted stellar
mass $M_* > 5 \times 10^5\, \Msun$ at $z=0$.  Such a cutoff agrees
with the observational limit of all Local Group dwarfs (Table
\ref{tab:sfdata_lg}) known until a few years ago, when the ultrafaint
dwarfs were discovered.  We defer the discussion of the model
predictions for these low-mass objects until \S\ref{sec:lowmass}.

As in KGK04, we do a rough morphological classification of dwarf
galaxies as dSph or dIrr based on the ratio of the stellar rotation
velocity to the velocity dispersion: $v_{\rm rot}/\sigma < 3$ for dSph
and $v_{\rm rot}/\sigma > 3$ for dIrr.  In the model, the rotation
velocity is calculated as the circular velocity at the outer-most
stellar radius, as it would be measured in observation, while the
velocity dispersion is estimated from the amount of external tidal
heating in strong tidal interactions with other halos.  Such
classification does not take into account the recent star formation
activity or the remaining gas content, and therefore, is only a crude
indication of the observationally defined morphological type.
Comparison of these model predictions to the data is, in fact,
completely complementary to the comparison of the star formation
histories.

In the process of revising our models we have discovered that the gas
densities $\Sigma_{g}$ in \citet{tumultuous04} were underestimated
by a factor of 2 due to an error in the code.  When we quote the
results for the KGK04 model here, we use the corrected values.

\subsection{Discrepancies of KGK04 Model with Star Formation Data}
  \label{sec:kgk}

Despite significant successes in explaining the number and spatial
distribution of the Local Group dwarfs, the KGK04 model did not
predict a sufficient amount of recent star formation.  Figure
\ref{fig:sfplots} shows that 95\% of the dwarfs in that model have not
formed any stars in the last 1 Gyr, in serious disagreement with data.

To quantify the level of this disagreement, we use the
Kolmogorov-Smirnov (KS) test for cumulative distribution functions of
the following parameters: the distance to the host ($r_{\rm all}$),
the distances for the dSph and dIrr galaxies separately ($r_{\rm
dSph}$ and $r_{\rm dIrr}$), the star formation fractions ($f_{1G}$,
$f_{2G}$, $f_{5G}$, $f_{10G}$), the mean age ($\age$), and the stellar
mass ($M_*$).  Table~\ref{tab:sfdata} shows that the recent star
formation fractions in the KGK04 model (see column $\epsilon_*=0$)
have very low KS probabilities, below $10^{-12}$.  On the other hand,
the $f_{10G}$ fraction, which is a more global measure of the overall
star formation, is consistent with the data at the 13\% level.  Thus
the star formation law used in the model is not necessarily at fault,
but the apparent lack of recent star formation is a direct result of
the fixed star formation threshold, as we will see below.

The mean mass-weighted age is also inconsistent with the data, but at
a less significant level ($P_{KS} \sim 10^{-3}$).  The mean age is a
global, integral measure of the SFH, which combines the
low-probability recent star formation with a higher-probability early
star formation.  The stellar age is systematically overestimated in
the model, by a few Gyr.

The total stellar mass is also overestimated by a factor of several
for most dwarfs.  The cumulative mass function is inconsistent with
the data at a level similar to the age distribution, $P_{KS} \sim
10^{-3}$.

Note also that KGK04 considered principally the satellite dwarfs located
within the virial radius of the host galaxy at $r < 200 \, h^{-1}$
kpc.  We extend our analysis to all dwarfs in the Local Group, even
the potentially isolated ones lying outside the virial radius of
either host.  Figure~\ref{fig:radial} shows their cumulative radial
distribution out to $1000 \, h^{-1}$ kpc.  The KGK04 model does well
for the distribution of all dwarfs and the distribution of Irr/dIrr
types separately.  The model slightly overestimates the number of
dSph/dE types outside $200 \, h^{-1}$ kpc, but in all cases the KS
probability is $\sim 10$\% or higher, fully consistent with the
observations.

Our phenomenological model contains several parameters that allow for
some freedom in the outcome. Two of the most 
important parameters in the model are the threshold density 
$\Sigma_{\rm th}$ and the disk size parameter, $c$ (eq. \ref{eq:r_d}),
both of which are difficult to know, {\it a priori}, from theory. 
Since these parameters can significantly change the predicted total 
number of dwarfs, the observed number of ($M_* > 5 \times 10^5 M_\odot$)
dwarfs provides a fairly stringent constraint on these values. For 
example, with $c=10$ and $\Sigma_{\rm th} = 5 \, \Mpc2$, the predicted 
average number of dwarfs per host halo within 1~$h^{-1}$~Mpc is 20.5.
If we take $c=5$ (with $\Sigma_{\rm th} = 5 \, \Mpc2$) this number 
increases to 40.7, and if we take $c=15$ the number of dwarfs drops to
14.3.  Analogously, reducing the threshold density to 4 and $3 \, \Mpc2$
(with $c=10$) increases the average number to 21.7 and 22.3, 
respectively.  None of these fixed threshold models, however, adequately
reproduces the observed star formation histories.  Therefore, we look for
additional physical ingredients for our model.

In our extension of the KGK04 model, we attempt to retain the correct
radial distribution of the dwarf galaxies, while improving the
predictions for their star formation histories.

\section{Stochastic Star Formation Threshold}
  \label{sec:threshold}

We consider a number of modifications to the KGK04 model. The most
promising of the modifications is the introduction of a stochastic
threshold to the star formation law.

At each output epoch through the course of the simulation, the
threshold density for star formation is drawn from a log-normal
distribution with a mean value $\Sigma_{\rm th0}$ and a small
dispersion $\epsilon_* \sim 0.1$:
\begin{equation}
  P(\Sigma_{\rm th}) \; d\Sigma_{\rm th} 
   = \frac{1}{\sqrt{2 \pi} \epsilon_*}
   \exp{\left[{-\frac{(\log{\Sigma_{\rm th}} - \log{\Sigma_{\rm th0}})^2}
              {2\epsilon_*^2}}\right]} \; d\Sigma_{\rm th}.
  \label{eq:sig_distrib}
\end{equation}
As a fiducial mean value we take $\Sigma_{\rm th0} = 5 \Mpc2$, but we
also vary that parameter in some runs.

What does this stochastic threshold mean?  And what, physically, is
the threshold of star formation?  We apply it to the
azimuthally-averaged surface density of gas.  In nearby star forming
regions, which we can study directly with {\it HST} and {\it Spitzer},
stars form in dense molecular clouds
\citep[e.g.,][]{mckee_ostriker07}.  Locally the density is high, but
the azimuthal average at a particular distance $r$ from the center may
be either high (if molecular clouds are common at $r$) or low (if they
are rare).  Thus, by invoking the threshold $\Sigma_{\rm th}$, we are
effectively parameterizing the fraction of molecular gas available for
star formation.  When we take the threshold to vary, we are thus
accounting for stochastic star formation in isolated HII regions, such
as those found by {\it GALEX} in nearby spirals
\citep{thilker_etal07}.  

Nearby dwarf galaxies often show a very high HI gas fraction, i.e. the
ratio of gas mass to baryon (gas+stars) mass, up to 90\%
\citep{fisher_tully1975, geha_etal06, lee_etal06}.  The current star
formation rates in those galaxies are low, and therefore most of the
gas is inert.  Hydrodynamic simulations of
\citet{RobertsonKravtsov2007}, which treat the formation of molecular
hydrogen in detail, also suggest that dwarf disks may contain large
reservoirs of diffuse atomic gas that is unable to condense in
molecular clouds and participate in star formation.  This inefficiency
of forming molecular clouds at low densities effectively results in
the threshold density for star formation.

\subsection{A Major Improvement}

Figure \ref{fig:sfplots} shows that the stochasticity greatly improves
the agreement of the recent star formation fractions ($f_{1G},
f_{2G}$, $f_{5G}$) with the Local Group data.  The predicted
distributions are much closer to the observed ones than in the KGK04
model, and are, in fact, statistically consistent with each other at
$\sim 10\%$ level.

We quantify this effect by gradually increasing the amount of
stochasticity, $\epsilon_*$.  Table~\ref{tab:sfdata} shows the KS test
results for $\epsilon_*$ ranging from 0.05 to 0.2, which approximately
corresponds to 10\% to 60\% variation in the threshold density
$\Sigma_{\rm th}$.  Even a small amount of stochasticity, $\epsilon_*
\sim 0.05$, leads to a dramatic improvement of the $f_{1G}, f_{2G}$,
and $f_{5G}$ statistics.  The probabilities increase rapidly with
$\epsilon_*$ to a good fraction of a percent or more.  The model with
$\epsilon_* = 0.1$ is already statistically consistent with the data.

With the other parameters of the star formation law ($\Sigma_{\rm
th0}$ and $\fsfr$) being fixed, increasing $\epsilon_*$ leads to
better recent star formation parameters but worse early star formation
parameters.  The probability of the $f_{10G}$ distribution decreases
from 13\% to under 5\% for $\epsilon_* = 0.1$ and even below a percent
for $\epsilon_* = 0.2$.

The mean age is most consistent with the data ($P_{KS} = 0.61$) for
$\epsilon_* = 0.1$.  The total number of dwarfs also increases
systematically with $\epsilon_*$ while the stellar mass function and
the radial distributions are not significantly affected by the
variation of $\epsilon_*$.  Additionally, the $\epsilon_* = 0.15$ and
$\epsilon_* = 0.2$ models begin to overpredict the total number of
luminous dwarfs per host halo, implying that at these values the
overall star formation becomes too efficient.  This overabundance can
be seen in Table~\ref{tab:sfdata}, or graphically in
Fig.~\ref{fig:radial}.  Therefore, we take the case with $\epsilon_* =
0.1$ as our fiducial model striking the best balance for all star
formation statistics.  We consider other variants of the stochastic
model in \S\ref{sec:other}.

\begin{figure*}
\vspace{-0.5cm}
\centerline{\epsfig{file=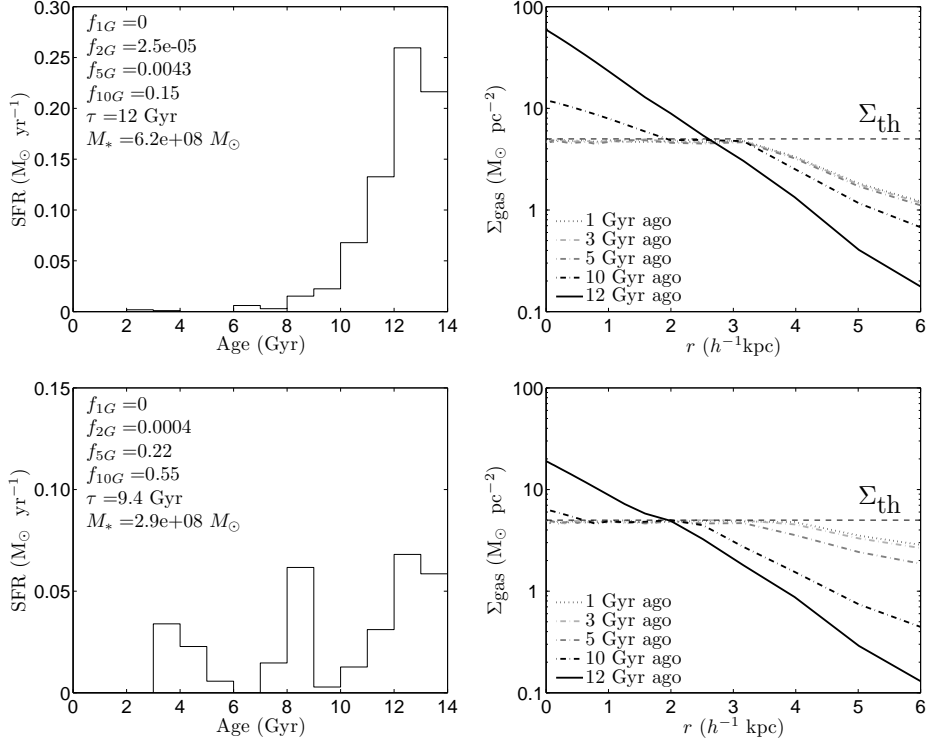, angle=0, width=5.75in}}
\vspace{-0.5cm}
\caption{{\it Left:} The star formation histories of two
simulated dwarfs in the KGK04 model, coarsened into 1 Gyr bins so as
to better resemble the time resolution of the data. The x-axis shows the 
Age (i.e. of the isochrone) with the present epoch at 0 Gyr, going back to 
the Big Bang at 14 Gyr. {\it Right:} The
gas density profiles for the two simulated galaxies that were used to
calculate their star formation rates, at five epochs.  The 
fixed star formation threshold is shown by the dashed
horizontal line.}
  \label{fig:sfr_gasprofile}
\end{figure*}

\begin{figure*}
\vspace{-0.5cm}
\centerline{\epsfig{file=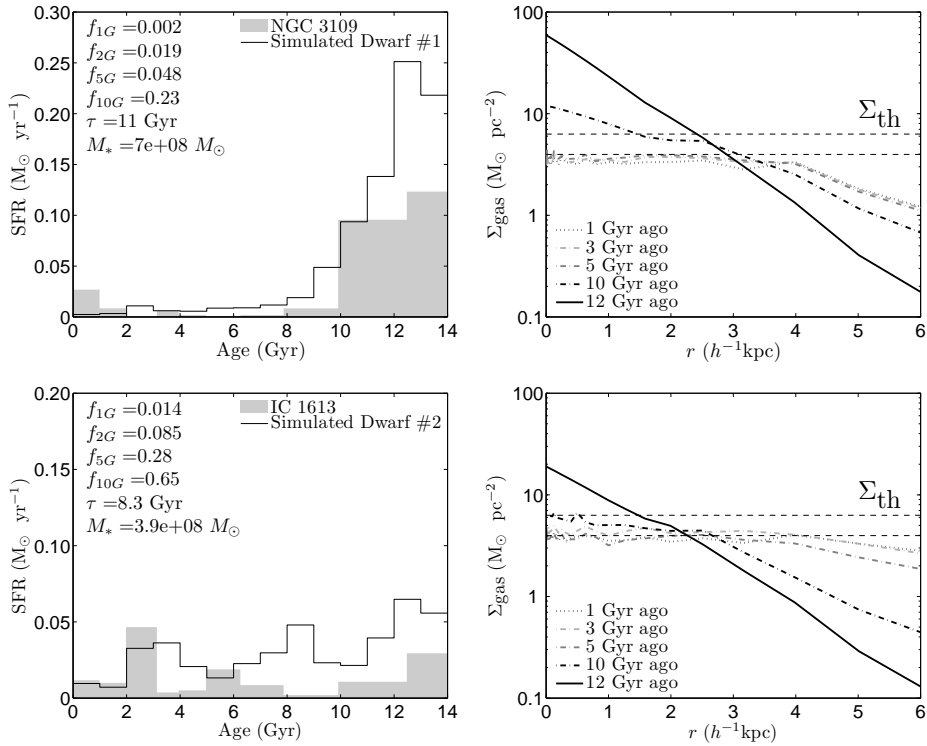, angle=0, width=5.75in}}
\vspace{-0.5cm}
\caption{{\it Left:} The star formation histories of the two dwarfs
from Fig.~\protect\ref{fig:sfr_gasprofile} in the model with a
stochastic star formation threshold, $\epsilon_* = 0.1$ (dashed
histograms and quoted parameters).  For comparison, shaded histograms
show the observed SFHs of NGC 3109 and IC 1613, normalized to the total
stellar mass cited in Table~\protect\ref{tab:sfdata_lg}.  {\it Right:}
The gas density profiles at five epochs.  The range of variation of
the star formation threshold is shown by dashed horizontal lines.
This range allows star formation to continue at late times.}
  \label{fig:sfr_vs_sfr}
\end{figure*}

\subsection{Reasons for Success}

Figures~\ref{fig:sfr_gasprofile} and \ref{fig:sfr_vs_sfr} illustrate
why the stochasticity is so successful.  They show star formation
histories and corresponding gas density profiles for two
representative model dwarfs.  Figure~\ref{fig:sfr_gasprofile} is for
the case of a fixed density threshold, $\Sigma_{\rm th} = 5 \Mpc2$.
Without the stochasticity, and in the absence of radial gas flows, the
gas above the threshold is steadily converted into stars at early
times (10 and 12 Gyr ago) at substantial rates, $0.1~-~0.3\ \Msun~\rm{yr}^{-1}$.  At these epochs the galaxies evolve effectively in
isolation, while still growing by gas-rich hierarchical mergers.  At
later time, when these galaxies become satellites of the larger host
galaxy, they are tidally truncated and no new gas is accreted.  After
the high-density gas supply is exhausted, the rest of the gas hovers
just under the threshold, unable to form new stars.  In the first
dwarf, shown in the top panels in Figure~\ref{fig:sfr_gasprofile}, 
star formation almost completely halts
8 Gyr ago.  The other dwarf, shown in the bottom panels, experiences
several distinct episodes of star formation, the last one finishing 3
Gyr ago.  In both cases there is effectively no star formation in the
last 2 Gyr, $f_{1G} \approx f_{2G} \approx 0$.

If instead the threshold density varies in time, at later epochs the
gas may find itself above the threshold and allow more recent star
formation.  Figure~\ref{fig:sfr_vs_sfr} shows the star formation
histories for the same dwarfs but now with a variable threshold and using
$\epsilon_* = 0.1$.  There is more star formation overall and more
star formation at later times.  As a result, the stellar mass is
higher by 20\% to 30\% and mean stellar age is lower by about 1 Gyr.
More importantly, several percent of all stars are formed in the last
2 Gyr.  This stochastic enhancement of the star formation rate allows
the model to reproduce the star formation episodes at late times when
gas-rich mergers (which increase the available gas supply) and strong
interactions (which can prompt starbursts) are relatively less common.

We also show on Fig.~\ref{fig:sfr_vs_sfr} the observed star formation
histories of two dwarf galaxies, NGC 3109 and IC 1613.  We do not expect
the model predictions to correspond in detail to the observed SFH
features, as our modeling is necessarily statistical and is aimed at
explaining not a specific galaxy's SFH but only an ensemble of SFHs of
many Local Group dwarfs.  Still, qualitatively, the agreement between
the model and the data is good: in the case of NGC 3109 both show most
stars being formed at early times with a small fraction in the last 4 Gyr,
while in the case of IC 1613, both SFHs continue until the present in
several distinct, extended episodes.

\subsection{Remaining Discrepancies}
  \label{sec:discrepancies}

Despite the impressive improvements, there still remain discrepancies
of the fiducial model with the observed data.  

First, there is still not quite enough very recent star formation, a
problem which is quantified by the KS-test result for $f_{1G}$ in
Table~\ref{tab:sfdata} ($P_{KS} \approx 3 \times 10^{-3}$) and is
apparent in Fig.~\ref{fig:sfplots}.  The models with a higher amount
of stochasticity, $\epsilon_* = 0.15$ and $\epsilon_* = 0.2$, achieve
better agreement with the observed $f_{1G}$ distribution but they are
disfavored for skewing all other star formation statistics.  Thus, in
the fiducial model ($\epsilon_* = 0.1$) over 25\% of the dwarfs have
less than $10^{-3}$ of their stars formed in the last 1 Gyr, compared
with about 10\% of such dwarfs in the observed sample.  Note that the
current observations are sensitive to SFHs with large fractions of
star formation at later times.  The very deep {\it HST} ACS imaging of Leo A
by \citet{cole_etal07} have confirmed that majority of all star
formation has occurred in the last half of the age of the universe.

Second, the total stellar masses of the luminous dwarfs in the models
tend to be significantly above the observed stellar masses of the
Local Group dwarfs.  This result is apparent for any value of
$\epsilon_*$ between 0 and 0.2 (see bottom right panel of
Fig.~\ref{fig:radial}).  This excess stellar mass in our models can
perhaps be traced to the fact that we do not include thermal and
ionizing feedback from young massive stars \citep[e.g.,][]{DekelSilk1986,
DekelWoo2003}, which can disrupt molecular clouds and drive galactic
outflows, thus reducing the available gas supply for star formation.

Third, the model does not predict enough very {\it early} star
formation.  This is a generic problem with any hierarchical model in
which the presently-massive satellites form and accrete onto the host
late and, at early times, therefore, do not contain significant
amounts of gas above the threshold.  Our model predicts that some 20\%
to 30\% of the dwarf galaxies will have formed greater than 85\% of
their stellar mass in the last 10 Gyr (i.e. $f_{10G} > 0.85$).  In
other words, there exists in the model a number of dwarfs with
anomalously young stellar populations, whereas, by contrast, the Local
Group data in Table~\ref{tab:sfdata_lg} does not show any dwarfs
having formed more than 86\% of its stellar mass in the last 10~Gyr
(the Saggitarius dSph has the highest fraction) -- at least 14\% of
all stars in the observed dwarfs formed in the first 4 Gyr after the
Big Bang.  Some of the simulated dwarfs with very young stellar
populations include objects which form the bulk of their stars in one,
punctuated star formation event caused by a close tidal interaction,
prompting a starburst which happened to occur in the last 10~Gyr.  But
not all of the galaxies in this category form stars through the
starburst mode; other dwarfs with very young stellar populations form
all of their stars through the continuous mode with no starbursts at
all.

\begin{table}[t]
\begin{center}
\caption{\sc Gas Accretion at $d < R_{\rm vir}$
   \label{tab:late_acc}}
\begin{tabular}{lcc}
\tableline\tableline\\
\multicolumn{1}{c}{} &
\multicolumn{1}{c}{No} &
\multicolumn{1}{c}{Yes}
\\[2mm] \tableline\\
$P_{KS}$: $r_{\rm all}$    & 2.5e-1  & 2.4e-1 \\           
$P_{KS}$: $r_{\rm dSph}$   & 3.8e-1  & 4.1e-1 \\           
$P_{KS}$: $r_{\rm dIrr}$   & 4.2e-2  & 3.1e-2 \\         
$P_{KS}$: $f_{1G}$         & 3.0e-3  & 9.1e-2 \\           
$P_{KS}$: $f_{2G}$         & 4.3e-1  & 8.1e-1 \\         
$P_{KS}$: $f_{5G}$         & 2.1e-2  & 1.9e-1 \\            
$P_{KS}$: $f_{10G}$        & 4.4e-2  & 5.8e-3 \\            
$P_{KS}$: $\age$           & 6.1e-1  & 3.8e-1 \\            
$P_{KS}$: $M_*$            & 3.4e-3  & 6.3e-4 \\            
Dwarfs per halo            & 22      & 23      \\            
\tableline
\end{tabular}
\end{center}
{\small {\sc Note.}---Other parameters are
$\epsilon_* = 0.1$, $\Sigma_{\rm th0} = 5 \Mpc2$, $\fsfr = 1$.}
\vspace{0.4cm}
\end{table}

\section{Other Variants of Star Formation Model}
  \label{sec:other}

With the discrepancies discussed in \S\ref{sec:discrepancies} in mind,
we have explored several variants of our star formation model in order
to check if relaxing other model assumptions can improve the predicted
star formation histories to the point where there is broad agreement
with the observations.

\subsection{Gas Accretion within the Virial Radius}

One of these alternate models is motivated by the lack of enough
recent star formation.  In the KGK04 model and in the stochastic
models shown in Figs.~\ref{fig:sfplots} and \ref{fig:radial},
accretion of new gas onto the dwarf halos is shut off whenever the
dwarf comes within the virial radius of the host halo.  This is based
on the expectation that satellite halos near their host would be
tidally truncated and unable to capture new gas even if they happen to
increase their dark matter mass in mergers with other satellite halos.
However, becoming a satellite does not immediately stop the halo's
star formation activity which can continue as long as the gas density
remains above the threshold or if a starburst event is triggered.

In the first of these alternative models, we lift this assumption and allow
the accretion of new gas even within the virial radius of the host, at
$d < R_{\rm vir}$.  As a result, the agreement between the model and
the data for the $f_{1G}$ distribution improves as indicated by the
KS-test results shown in Table~\ref{tab:late_acc} -- from the
rejection of the null hypothesis of the same-distribution at 0.3\%
significance level to rejection only at the 9\% level, essentially
statistically consistent.

Qualitatively, the improvement is most dramatic for $f_{1G}$ since
allowing the additional accretion of gas is more important at late
times, and because the virial radius of the host halo will be larger
than at earlier epochs; also the dwarf galaxies at late times are
likely to be closer in.  The extra gas is converted into extra stars
and so the $f_{1G}$ value increases and the typical mass of the dwarfs
increases.  Generally, the extra accretion shifts the stellar age
distribution to be younger.  However, another predicted feature of our models
 is a tail of the age distribution at $\age~<~5$~Gyr.  The extra accretion
exaggerates this tail since it increases recent star formation.  In
contrast, the youngest dwarf galaxy in the Local Group is Leo A, with
$\age = 6.2$ Gyr (though see the Leo A SFH of \citealt{cole_etal07},
which has a significantly younger result for $\age$). This is
fundamentally the same problem as the existence of simulated dwarfs
with $f_{10G} > 0.85$, and the $f_{10G}$ distribution similarly
becomes less consistent with the data in this model.

\begin{table}[t]
\begin{center}
\caption{\sc Slope of Schmidt law, $\dot{\Sigma}_{*} \propto \Sigma_{g}^n$
   \label{tab:powerlaw}}
\begin{tabular}{lcccccc}
\tableline\tableline\\
\multicolumn{1}{c}{} &
\multicolumn{1}{c}{$n = 1.4$} &
\multicolumn{1}{c}{$n = 1.4$} &
\multicolumn{1}{c}{$n = 2$} &
\multicolumn{1}{c}{$n = 3$} 
\\[2mm] \tableline\\
$\fsfr$                    & 1       & 0.5     & 0.14    & 0.012  \\
$P_{KS}$: $r_{\rm all}$    & 2.5e-1  & 2.5e-1  & 2.6e-1  & 2.6e-1 \\
$P_{KS}$: $r_{\rm dSph}$   & 3.8e-1  & 4.4e-1  & 4.2e-1  & 4.8e-1 \\
$P_{KS}$: $r_{\rm dIrr}$   & 4.2e-2  & 3.5e-2  & 3.3e-2  & 3.5e-2 \\
$P_{KS}$: $f_{1G}$         & 3.0e-3  & 1.4e-2  & 2.5e-2  & 8.3e-2 \\
$P_{KS}$: $f_{2G}$         & 4.3e-1  & 2.4e-1  & 2.4e-1  & 1.8e-1 \\
$P_{KS}$: $f_{5G}$         & 2.1e-2  & 1.1e-1  & 9.4e-2  & 1.3e-1 \\ 
$P_{KS}$: $f_{10G}$        & 4.4e-2  & 5.7e-3  & 7.7e-3  & 8.3e-3 \\
$P_{KS}$: $\age$           & 6.1e-1  & 1.1e-1  & 4.5e-1  & 5.2e-1 \\
$P_{KS}$: $M_*$            & 3.4e-3  & 3.7e-3  & 3.1e-3  & 5.0e-3 \\
Dwarfs per halo            & 22      & 22      & 22      & 21     \\ 
\tableline
\end{tabular}
\end{center}
{\small
{\sc Note.}---Other parameters are
$\epsilon_* = 0.1$, $\Sigma_{\rm th0} = 5 \Mpc2$.}
\vspace{0.4cm}
\end{table}

\subsection{Modifications of the Schmidt Law}

Another important question to be addressed is whether the observed
SFHs, given our model assumptions, favor steepening of the
Schmidt-Kennicutt law ($\dot{\Sigma}_{*} \propto \Sigma_g^n$).  There
is some evidence for the exponents $n$ to become gradually larger at
low gas densities in dwarf galaxies (e.g., \citealt{boissier_etal03,
heyer_etal04}; see also discussion in
\citealt{RobertsonKravtsov2007}).  Since most of our dwarfs are close
to the threshold, we investigate this effect simply by changing the
exponent at all densities, setting it to $n = 1.4$, 2, or 3.  For each
of these values, the normalization $\fsfr$ is allowed to float such
that the overall star formation efficiency is lowered until the point
where the median of the age distribution, $\age_{\rm med}$, of the
simulated dwarfs matches that of the Local Group galaxies, 9.25 Gyr.

Table~\ref{tab:powerlaw} shows the required values of $\fsfr$ and the
results of KS-tests.  The rather high significance results from the KS
tests applied to the $\age$-distribution are not particularly
surprising since the normalization of the star formation law has been
fine-tuned for this comparison.  Qualitatively this exercise keeps the
overall star formation rate the same, i.e.  the stellar mass function
is basically unchanged.  And similarly the radial distributions stay
roughly the same -- this is important since one of the primary
successes of the KGK04 model was reproducing the observed radial
distribution of the Local Group dwarfs, something that our models
continue to do.  Eventually the same amount of gas above the density
threshold is converted into stars, so the rate at which star formation
proceeds does not alter the total number or the radial distribution of
the simulated dwarfs.

By and large, all of the same discrepancies with the data listed in
\S\ref{sec:discrepancies} are apparent in these models with different
$n$ -- too large stellar masses, not enough recent star formation, and
too much overall star formation in the last 10 Gyr.  Arguably the \{$n
= 3$, $\fsfr = 0.012$\} model is a best fit to the data for $f_{1G}$
and $f_{5G}$, however the fiducial \{$n = 1.4$, $\fsfr = 1$\} model is
still the best for $f_{2G}$ and $f_{10G}$.  Further, comparisons of
the stellar mass and the $\tau$-distribution are inconclusive as well.

\begin{table}
\begin{center}
\caption{\sc Reionization Scenarios 
   \label{tab:reioniz}}
\begin{tabular}{lcccccc}
\tableline\tableline\\
\multicolumn{1}{c}{} &
\multicolumn{1}{c}{Early} &
\multicolumn{1}{c}{Extended} &
\multicolumn{1}{c}{Fiducial} &
\multicolumn{1}{c}{Late}
\\[2mm] \tableline\\
$z_r$                      & 9        & 7       & 7        & 6      \\
$z_o$                      & 10       & 10      & 8        & 10     \\
$P_{KS}$: $r_{\rm all}$    & 2.3e-1   & 2.9e-1  & 2.5e-1   & 3.2e-1 \\
$P_{KS}$: $r_{\rm dSph}$   & 5.7e-1   & 5.6e-1  & 3.8e-1   & 5.3e-1 \\
$P_{KS}$: $r_{\rm dIrr}$   & 4.2e-2   & 4.7e-2  & 4.2e-2   & 6.0e-2 \\
$P_{KS}$: $f_{1G}$         & 1.6e-2   & 7.7e-3  & 3.0e-3   & 1.3e-2 \\
$P_{KS}$: $f_{2G}$         & 7.7e-1   & 6.4e-1  & 4.3e-1   & 6.6e-1 \\
$P_{KS}$: $f_{5G}$         & 1.2e-1   & 8.2e-2  & 2.1e-2   & 6.3e-2 \\ 
$P_{KS}$: $f_{10G}$        & 4.8e-3   & 9.9e-3  & 4.4e-2   & 1.5e-2 \\
$P_{KS}$: $\age$           & 6.3e-1   & 6.9e-1  & 6.1e-1   & 6.7e-1 \\
$P_{KS}$: $M_*$            & 1.6e-3   & 1.3e-3  & 3.4e-3   & 2.4e-3 \\
Dwarfs per halo            & 20       & 19      & 22       & 18      \\ 
\tableline
\end{tabular}
\end{center}
{\small
{\sc Notes.}---$z_r$ refers to the redshift when reionization is completed,
while $z_o$ refers to the redshift where cosmic HII regions begin to
overlap (see Appendix B in \citealt{tumultuous04} for details).
Other parameters:
$\epsilon_* = 0.1$, $\Sigma_{\rm th0} = 5 \Mpc2$, $\fsfr = 1$.}
\vspace{0.4cm}
\end{table}

\subsection{Revised Epoch of Reionization and Cutoff Mass}

Another interesting test is to vary the epoch of reionization.
Appendix B of KGK04 presents an analytic fit to the numerical results
of \citet{Gnedin2000} that parametrize the reionization epoch in terms
of the redshift when cosmic HII regions begin to overlap, $z_o$, and
the redshift when the reionization is completed, $z_r$.  Varying these
two parameters allows us to test the sensitivity of model predictions
to the details of the evolution of the ionizing background radiation.
Note that our parametrization of reionization here applies only to the
local neighborhood of the Galaxy and may differ from the global cosmic
reionization.

Reionization affects the amount of gas accreted onto dwarf halos.
Increasing extragalactic UV flux during reionization photoionizes the
gas inside and outside dark matter halos and prevents the halos with
shallow potential wells from capturing new gas heated to $\sim 10^4$
K.  In \citet{Gnedin2000} the effect of decreasing the gas fraction of
halos was parametrized as 
\begin{equation}
  f_{\rm gas} = f_{\rm b}\; (1 + M_c/M)^{-3},
  \label{eq:fgas}
\end{equation}
where $f_{\rm b}$ is the universal baryon fraction, $M$ is the halo
mass, and $M_c$ is the cut-off mass parameter.  In linear theory for
baryon perturbations, this parameter can be related to the filtering
mass as $M_c \approx 0.26 \, M_f$.  The filtering mass $M_f$ is an
integral of a function of the intergalactic gas temperature over
cosmic history and corresponds to the mass of a halo that loses 50\%
of its baryons as a result of external photoheating.  The values of
the filtering mass from \citet{Gnedin2000} can be calculated for any
$z_o$ and $z_r$ using eq. (B1) in \citet{tumultuous04}.

We consider four scenarios listed in Table~\ref{tab:reioniz}: early
\{$z_r = 9$, $z_o = 10$\}, extended \{$z_r = 7$, $z_o = 10$\}, and late
\{$z_r = 6$, $z_o = 10$\} reionizations, in addition to the standard
scenario \{$z_r = 7$, $z_o = 8$\}, which was used in KGK04 and in our
fiducial model.  Generally, the results are quite similar for all
scenarios, likely because luminous satellites are hosted by the
relatively massive halos, in which the virial temperatures are above
$10^4$ K and the external heating of the gas does not affect its
distribution significantly.  Early reionization is slightly preferred
for the $f_{1G}$, $f_{2G}$, and $f_{5G}$ distributions, but
statistically all scenarios are consistent with each other.  The
$f_{10G}$ distribution is still reproduced best by the fiducial model.

Recent hydrodynamic simulations indicate that the filtering mass may
overestimate the mass of the halos that lose 50\% of their baryons,
especially at low redshift.  The most sophisticated ART simulations of
\citet{tassis_etal08}, including the effects of radiative transfer,
give distributions of $f_{\rm gas}$ at several redshifts, $z = 3.3, 4,
5, 7, 8, 9$ (their Fig.~2) for the standard model of reionization.  We
have fit these distributions with the form of equation (\ref{eq:fgas})
and obtained best-fitting values of $M_c$.  These values are similar
to our old values at $z \ge 8$ but deviate from them at lower redshift
by as much as an order of magnitude.  We supplement these
high-redshift fits with the $z=0$ results of SPH simulations by
\citet{Crain_et_al_2007} and \citet{Hoeft_et_al_2007}, which both give
$M_c \sim 2\times 10^9\, h^{-1}\ \Msun$, about a factor of 5 smaller
than our old value.  The combined sets at $z=0$ and $z>3$ can be fit
by the following expression, accurate to better than 50\%:
\begin{equation}
  \label{eq:new_mc}
  M_c \approx 1.8 \times 10^7 + 3 \times 10^9 \, (1+z)^{-3} \ h^{-1}\, \Msun.
\end{equation}
Ideally, we would like to derive the cutoff mass evolution from
several simulations covering the whole range of redshifts, but at the
moment it is the best we can assemble from the literature.

We have run our model with the new expression (\ref{eq:new_mc}) for
$M_c$ and found that the corresponding changes in the predicted SFHs
are not straightforwardly better or worse.  The $f_{1G}$
distribution improves a little, $f_{2G}$ and $f_{5G}$ are effectively
unchanged, while the $f_{10G}$ distribution is significantly worse off 
 and the anomalously young dwarfs still persist.  
The age distribution and
stellar mass function are also more discrepant with the data than in
the fiducial model.  While the lack of improvement with the new cutoff
mass prescription is unpleasant or, at any rate, suprising, until we 
have a more robust estimate of $M_c$ confirmed by several groups, 
we keep our fiducial model untouched.

\begin{figure*}
\vspace{-1.0cm} 
\centerline{\epsfig{file=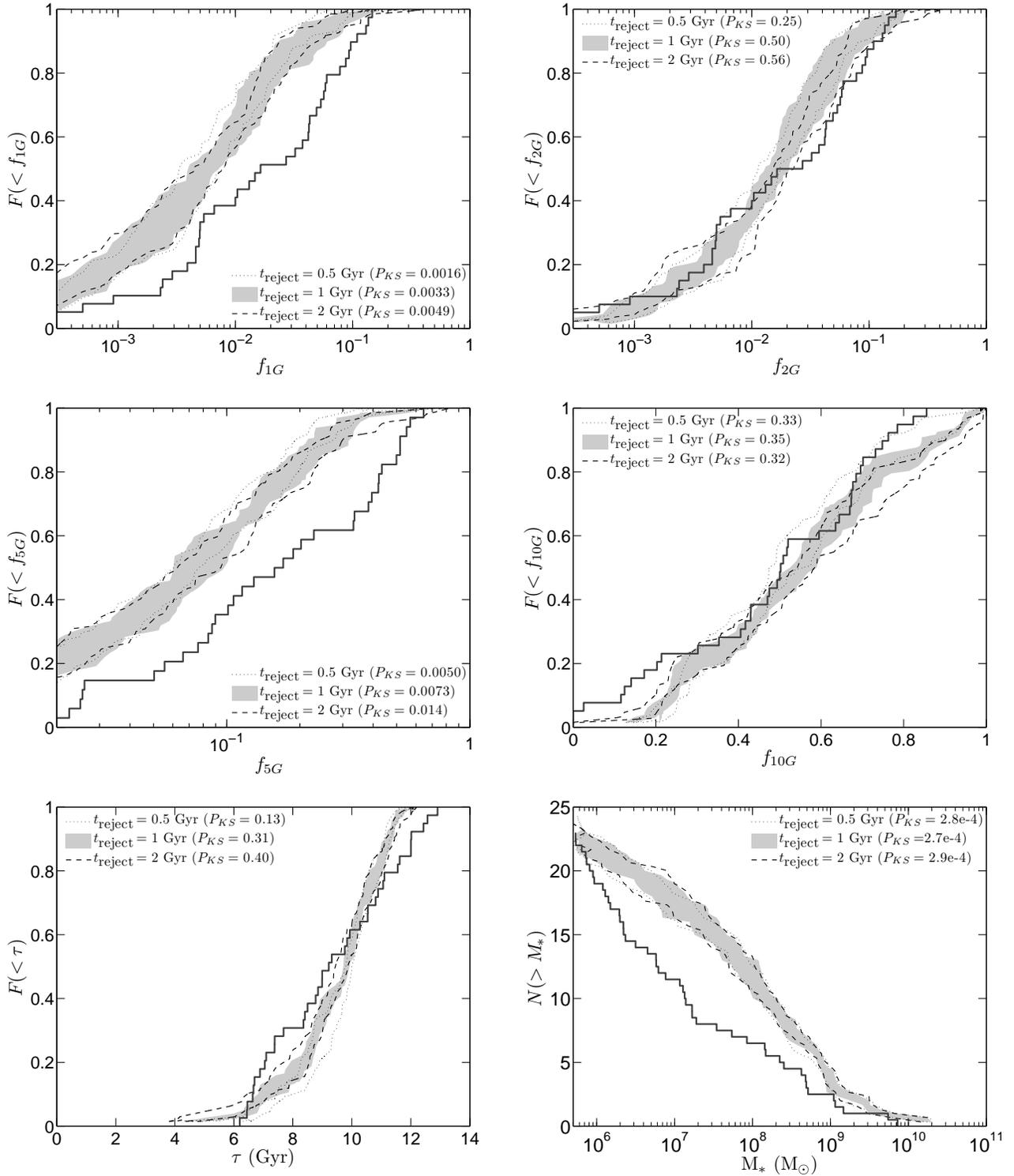, angle=0,width=8.0in}}
\vspace{-1.6cm}
\caption{ A comparison of three variants of the fiducial model which
reject dwarfs that have not formed any stars by $t_{\rm
reject}~=$~0.5~Gyr (filled gray shape), $t_{\rm reject}~=$~1.0~Gyr
(dashed line), or $t_{\rm reject}~=$~2.0~Gyr (dotted line) after the
Big Bang.  Local Group data are shown as a solid black line.  Note
that in order to match the observed number of dwarfs, the star
formation density threshold is lowered in these models, see
Table~\ref{tab:reject_LB2}.  The only noticeable improvements,
relative to the fiducial model, are in the $f_{10G}$ distribution and
in the corresponding lack of the low-$\age$ tail in the age
distribution.}
  \label{fig:sfplots_early2}
\end{figure*}

\begin{table}[b]
\begin{center}
\caption{\sc Stochastic vs. Monotonic Threshold \\ of Star Formation Law
   \label{tab:mono}}
\begin{tabular}{lccc}
\tableline\tableline\\
\multicolumn{1}{c}{} &
\multicolumn{1}{c}{$\epsilon_* = 0$} &
\multicolumn{1}{c}{$\epsilon_* = 0.1$} &
\multicolumn{1}{c}{$\epsilon_* = 0$}  \\
\multicolumn{1}{c}{} &
\multicolumn{1}{c}{$\Sigma_{\rm th0} = 5$} &
\multicolumn{1}{c}{$\Sigma_{\rm th0} = 5$} &
\multicolumn{1}{c}{$\Sigma_{\rm th}(z)$} 
\\[2mm] \tableline\\
$P_{KS}$: $r_{\rm all}$    & 3.1e-1  & 2.5e-1  & 3.0e-1 \\
$P_{KS}$: $r_{\rm dSph}$   & 3.7e-1  & 3.8e-1  & 3.4e-1 \\
$P_{KS}$: $r_{\rm dIrr}$   & 8.7e-2  & 4.2e-2  & 3.7e-2 \\
$P_{KS}$: $f_{1G}$         & 1.1e-19 & 3.0e-3  & 3.7e-2 \\
$P_{KS}$: $f_{2G}$         & 2.0e-16 & 4.3e-1  & 7.1e-1 \\
$P_{KS}$: $f_{5G}$         & 3.8e-13 & 2.1e-2  & 1.6e-2 \\
$P_{KS}$: $f_{10G}$        & 1.3e-1  & 4.4e-2  & 2.7e-2 \\
$P_{KS}$: $\age$           & 1.7e-3  & 6.1e-1  & 5.0e-1 \\
$P_{KS}$: $M_*$            & 6.2e-3  & 3.4e-3  & 2.0e-3 \\
Dwarfs per halo            & 20      & 22      & 23     \\
\tableline
\end{tabular}
\end{center}
\end{table}

\subsection{Monotonically Variable Star Formation Threshold}
 \label{sec:mono_sec}
                                                                              
In addition to exploring the stochastically variable density threshold,
we have also investigated a threshold $\Sigma_{\rm th}$ that varies
with redshift monotonically.  Though, at the moment, we do not know of a
convincing physical motivation for such a global systematic variation, 
we have investigated this possibility in exploring the full range of
predictions of our model.  Here we used a simple parameterization of
 the $\Sigma_{\rm th}$ redshift dependence,
\begin{equation}
  \Sigma_{\rm th}(z) = \Sigma_{\rm th0} \, (1+z)^\alpha,
  \label{eq:sig_mono}
\end{equation}
where $\Sigma_{\rm th}(z=0) = \Sigma_{\rm th0} = 3 \, \Mpc2$
and $\Sigma_{\rm th}(z=9) = 5 \, \Mpc2$, which results in $\alpha =
\log{(5/3)} \approx 0.22$.
                                                                               
Interestingly, the results of this model are very similar to our fiducial
model with the stochastic threshold.  The comparison is shown in 
Table~\ref{tab:mono}.  The new model shares the same problems as the
fiducial model: the stellar masses are still too large and 
the anomalously young dwarf problem is still present (if not slightly
worse). We do not consider any other variants of the monotonic threshold 
but note that if a compelling theoretical or observational  motivation 
for such a variation appears in the future, an evolving threshold may 
become a viable model.

\begin{table}
\begin{center}
\caption{\sc Rejecting Late Beginners
   \label{tab:reject_LB2}}
\begin{tabular}{lcccccc}
\tableline\tableline\\
\multicolumn{1}{c}{} &
\multicolumn{1}{c}{$t_{\rm reject} =$} &
\multicolumn{1}{c}{$t_{\rm reject} =$} &
\multicolumn{1}{c}{$t_{\rm reject} =$} &
\multicolumn{1}{c}{No} \\
\multicolumn{1}{c}{} &
\multicolumn{1}{c}{0.5 Gyr} &
\multicolumn{1}{c}{1 Gyr} &
\multicolumn{1}{c}{2 Gyr} &
\multicolumn{1}{c}{rejection}
\\[2mm] \tableline\\
$\Sigma_{\rm th0}$         & 1.87    & 2.6    & 3.5     & 5.0    \\
$P_{KS}$: $r_{\rm all}$    & 5.3e-1  & 4.0e-1 & 3.4e-1  & 2.5e-1 \\
$P_{KS}$: $r_{\rm dSph}$   & 1.7e-1  & 3.7e-1 & 5.1e-1  & 3.8e-1 \\
$P_{KS}$: $r_{\rm dIrr}$   & 2.2e-1  & 9.6e-2 & 7.3e-2  & 4.2e-2 \\
$P_{KS}$: $f_{1G}$         & 1.6e-3  & 3.3e-3 & 4.9e-3  & 3.0e-3 \\
$P_{KS}$: $f_{2G}$         & 2.5e-1  & 5.0e-1 & 5.6e-1  & 4.3e-1 \\
$P_{KS}$: $f_{5G}$         & 5.0e-3  & 7.3e-3 & 1.4e-2  & 2.1e-2 \\ 
$P_{KS}$: $f_{10G}$        & 3.3e-1  & 3.5e-1 & 3.2e-1  & 4.4e-2 \\
$P_{KS}$: $\age$           & 1.3e-1  & 3.1e-1 & 4.0e-1  & 6.1e-1 \\
$P_{KS}$: $M_*$            & 2.8e-4  & 2.7e-4 & 2.9e-4  & 3.4e-3 \\
Dwarfs per halo            & 22      & 22     & 23      & 22     \\ 
\tableline
\end{tabular}
\end{center}
{\small {\sc Notes.}---Here we reject dwarfs that have not formed any
stars by $t_{\rm reject} = 0.5$ Gyr, 1 Gyr, or 2 Gyr after the Big
Bang (redshifts $z \approx 10$, $z \approx6$, and $z \approx 3.3$,
respectively).  Other parameters: $\epsilon_* = 0.1$, $\fsfr = 1$.}
\vspace{0.4cm}
\end{table}

\subsection{Rejecting Galaxies with Delayed Star Formation}

Finally, in directly addressing the problematic issue of the simulated
dwarfs with too young stellar populations, we consider models that
reject dwarfs with delayed star formation, i.e. the dwarfs that have
not formed any stars within the first few Gyr after the Big Bang.

The reasoning for such an {\it ad hoc} cut is motivated by the
uncertainty in the detailed effect of the early UV background on the
gas content of low-mass halos.  The satellites without early star
formation in our model must have acquired a significant gas reservoir
only at late times.  In the hierarchical paradigm, these satellites
have been built by mergers of smaller objects, each of which carried
an even smaller amount of gas.  In order to become a galaxy in our
model, at some time the total gas density distribution in the
satellite must reach above the density threshold.  However, if for any
reason we underestimated the effect of gas loss from small halos at
early times, then at that time the combined amount of gas would be
overestimated and star formation should not take place.  Our treatment
of the gas heating during reionization is very approximate, which
makes it likely that we could either underestimate or overestimate the
gas loss effect for particular dwarfs.  In general, star formation at
very high redshift could proceed even at lower densities than assumed
in our models, such that all luminous dwarfs form at least a fraction
of their stars before reionization \citep{ricotti_gnedin05}.

In the current concordance cosmology \citep{wmap5}, the epoch of
complete reionization can range from $z \sim 10$ to $z \sim 6$, which
corresponds to a range of times from 0.5 to 1 Gyr after the Big Bang.
Given this uncertainty, and uncertainties in the details of gas outflows, we
consider three models that reject the dwarfs without star formation in
the first 0.5 Gyr, 1 Gyr, or 2 Gyr after the Big Bang.  The last model
with $t_{\rm reject} = 2$ Gyr is the least restrictive and closest to
the fiducial model (no rejection), which can be formally written as
$t_{\rm reject}~>~14~\rm{Gyr}$.  In each of these ``reject late beginners''
models, the star formation density threshold $\Sigma_{\rm th0}$ is
lowered in order to raise the number of dwarfs back to the number
observed in the Local Group.  The details of these models are listed
in Table~\ref{tab:reject_LB2}.

\begin{figure}[t]
\centerline{\epsfig{file=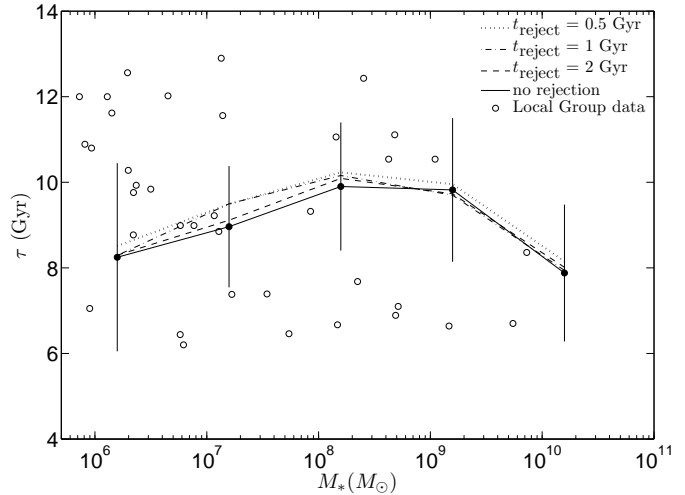, angle=0, width=4.0in}}
\caption{Mass-weighted mean stellar age vs. stellar mass at $z = 0$.
We show the fiducial model and three variants, which reject simulated
galaxies with delayed star formation, by 0.5, 1, and 2 Gyr,
respectively.  The model galaxies are binned by stellar mass and only
the bin averages are shown for clarity.  Vertical bars indicate the
standard deviation of the sample in each bin.  Observed values for the
Local Group (Table~\protect\ref{tab:sfdata_lg}) are plotted by circles
and fall in the same range as the model.}
  \label{fig:noearly_t_vs_M}
\end{figure}

Figure~\ref{fig:sfplots_early2} shows the star formation parameters in
the three models.  All of the distributions are similar to each other,
but they all deviate from the fiducial model (plotted in corresponding
panels of Fig.~\ref{fig:sfplots}) in one important aspect.  The
$f_{10G}$-distribution lacks objects with $f_{10G}=1$, i.e. with no
star formation in the first 4 Gyr after the Big Bang.  Unfortunately,
the time resolution of the observed SFHs does not allow us to
discriminate among the three variants of the cut.  But all of them
provide the needed fix: the $f_{10G}$-distribution is fully consistent
with the data ($P_{KS} > 30\%$).

Table~\ref{tab:reject_LB2} shows that the other probabilities remain
roughly the same or even decrease relative to the fiducial model.  If
we are to accept any of the ``reject late beginners'' variants, we
would prefer the least restrictive $t_{\rm reject} = 2$ Gyr model.
Ideally, of course, we prefer to develop a better understanding of
early star formation that would make this cut unnecessary.

Figure~\ref{fig:noearly_t_vs_M} shows the stellar age as a function of
stellar mass in the fiducial model and its three variants.  Rejecting
any of the ``late beginners'' increases the stellar age by less than 1
Gyr, which is significantly smaller than the dispersion of the sample
at all masses.  A general trend, also largely overcome by the
dispersion, is for the mean age to increase with mass until $M_* \sim
10^9\, \Msun$ and then to decrease at larger masses.  The Local Group
data are consistent with the younger stellar ages, and extended SFHs,
for more massive galaxies.  At the lowest-mass end, however, the
observations show a number of very old objects that are still not
present in our models, even with the strictest age cut.  It is
apparent from this and previous plots that our prescription for star
formation in the first few Gyr of cosmic time still needs
improvement.

It is possible that the starburst mode of star formation is causing
very young stellar ages in small galaxies.  We have checked, however,
that the anomalously young dwarfs do not all have significant
starbursts at late times and are hosted by dark matter halos with a
wide range of masses.  In fact, galaxies with the highest fraction of
stellar mass built in starbursts are typically old, with $\tau$
between 10 and 12 Gyr.  Starbursts are not very important overall --
91\% of the dwarfs have less than 10\% of their stellar mass formed in
the starburst mode.  Also, we have varied the two parameters
describing the starburst mode (the minimum required tidal force and
the fraction of gas converted into stars) and found that it has little
effect on the SFH distributions.

It should also be mentioned that there does seem to be a trend, at $z
= 0$, for the anomalously young dwarf halos to be more massive than
the other, more typical, dwarf halos in the model -- observationally
speaking, these dwarfs would have higher dynamical mass-to-light
ratios.  Interestingly, at the end of reionization these anomalous
dwarfs are hosted by halos with a very wide range of masses, so that
there is significant dynamical evolution from the epoch of
reionization until the present day.  In other words, the younger
dwarfs really do have rather ``tumultuous'' lives.

\begin{table}
\begin{center}
\caption{\sc Low Mass Satellite Galaxies of MW and M31
  \label{tab:lowmass_tbl}}
\begin{tabular}{lcrccc}
\tableline\tableline\\
\multicolumn{1}{l}{Galaxy} &
\multicolumn{1}{c}{Host} &
\multicolumn{1}{c}{$r_{\rm host}$ (kpc)} &
\multicolumn{1}{c}{$M_{V}$} &
\multicolumn{1}{c}{$M_*$ ($M_{\sun})$}&
\multicolumn{1}{c}{Refs.} 
\\[2mm] \tableline\\
Leo T             & MW  & 417 & $-8.0$ & $4.0 \times 10^5$ &  1 \\
Canes Venatici I  & MW  & 218 & $-7.9$ & $3.7 \times 10^5$ &  2 \\
AndXI             & M31 & 103 & $-7.3$ & $2.1 \times 10^5$ &  3 \\
AndXIII           & M31 &  95 & $-6.9$ & $1.5 \times 10^5$ &  3 \\
AndXII            & M31 & 116 & $-6.4$ & $9.3 \times 10^4$ &  3 \\
Hercules          & MW  & 138 & $-6.0$ & $6.4 \times 10^4$ &  4 \\
Bo\"{o}tesI       & MW  &  62 & $-5.8$ & $5.4 \times 10^4$ &  5 \\
Ursa Major I      & MW  & 106 & $-5.6$ & $4.5 \times 10^4$ &  6 \\
LeoIV             & MW  & 158 & $-5.1$ & $2.8 \times 10^4$ &  4 \\
Canes Venatici II & MW  & 151 & $-4.8$ & $2.1 \times 10^4$ &  4 \\ 
SDSSJ100+5730     & MW  &  83 & $-4.2$ & $1.2 \times 10^4$ &  7 \\
SDSSJ1329+2841    & MW  &  76 & $-3.9$ & $9.4 \times 10^3$ &  7 \\
Ursa Major II     & MW  &  32 & $-3.8$ & $8.5 \times 10^3$ &  8 \\          
Coma Berencies    & MW  &  44 & $-3.7$ & $7.8 \times 10^3$ &  4 \\         
Bo\"{o}tesII      & MW  &  60 & $-3.1$ & $4.5 \times 10^3$ &  9 \\
WillmanI          & MW  &  38 & $-2.5$ & $2.6 \times 10^3$ &  10 \\
\tableline
\end{tabular}
\end{center}
{\small {\sc References.}---(1) \citealt{deJong_et_al_2008}; (2)
\citealt{Zucker2006}, \citealt{martin_etal08};  
(3) \citealt{Martin2006}; (4) \citealt{Belokurov2007}; 
(5) \citealt{Belokurov2006}, \citealt{siegel06}; 
(6) \citealt{SimonGeha2007}; (7) \citealt{Liu_etal08};
(8) \citealt{Zucker2006b}; (9) \citealt{Walsh2007};
(10) \citealt{Willman2006}, \citealt{siegel06}.\\
{\sc Notes.}---$M_*$ is estimated assuming $M_*/L_V = 3\, \Msun/\Lsun$, with
$L_V$ determined from $M_V$ quoted in the references.  Satellites of
MW are discovered by the SDSS, satellites of M31 by the MegaCam
survey. Satellite-to-host radii, $r_{\rm host}$, for AndXI, AndXII, \&
AndXIII are derived using their projected distances from M31, assuming
$d_{\rm M31} = 785$ kpc \citep{McConnachie_etal2005}.}
\end{table}

\begin{figure}
\vspace{-0.2cm}
\centerline{\epsfig{file=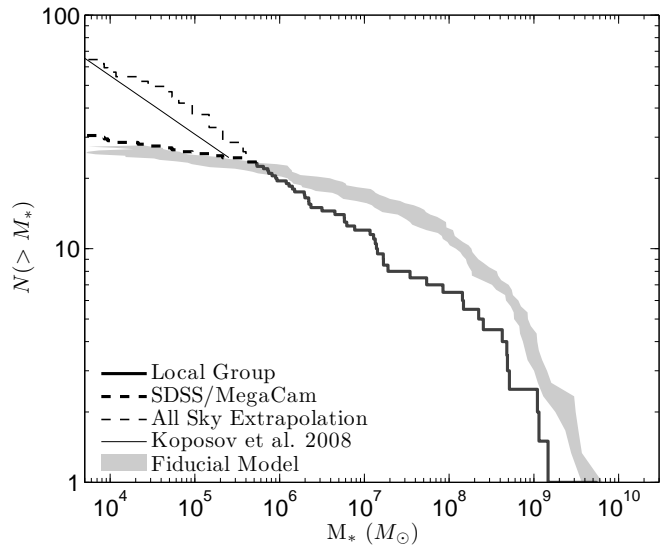, angle=0, width=4.0in}}
\caption{The stellar mass function per host halo, extended down to $5
\times 10^3\, \Msun$.  All dwarfs in our fiducial model at $d < 1\,
h^{-1}$ Mpc are shown in gray.  The Local Group data for $M_* > 5
\times 10^5\, \Msun$ are shown with a thick solid line, while the
thick dashed line shows the number of presently-known low mass ($M_* <
5 \times 10^5\, \Msun$) dwarfs (all the SDSS and MegaCam dwarfs listed
in Table~\ref{tab:lowmass_tbl}).  The thin dashed line is a likely
extrapolation of the mass function to all sky, to account for the
incompleteness of current surveys.  Thin solid line is an estimate of
the luminosity function by \citet{koposov_etal08}, assuming $M_* / L_V
= 3$.}
  \label{fig:lowmass_projections}
\end{figure}

\section{Projections for Low-Mass Dwarfs}
  \label{sec:lowmass}

Given the amazing rate of recent discoveries of the population of
ultrafaint dwarfs in the SDSS and MegaCam surveys, there stands a
challenge to predict the yet-to-be-observed star formation properties
of these objects.  Our models should in principle be able to predict
the mean age and stellar masses for the low-mass dwarfs, however
current predictions are not satisfactory.

Figure~\ref{fig:lowmass_projections} shows our fiducial model (with
the stochasticity parameter $\epsilon_* = 0.1$) extended to masses as
low as $M_* = 5 \times 10^3\ \Msun$.  This model reproduces most
closely the observed SFHs of the higher-mass dwarfs, but at $M_* < 5
\times 10^5 \ M_{\sun}$, it predicts only a modest increase in the
number of galaxies.  The thick dashed line in
Fig.~\ref{fig:lowmass_projections} shows the currently-known number of
ultrafaint dwarfs, listed in Table~\ref{tab:lowmass_tbl}.  The model
predictions actually agree very well with the observed number and with
the gentle slope of the mass function.

The problem, though, is that the SDSS data release 5, where the bulk
of the new dwarfs have been discovered, covers only 20\% of the sky.
Therefore, we might expect the full sky to contain 5 times as many
yet-to-be-discovered ultrafaint dwarfs \citep{SimonGeha2007}.  For the
MegaCam survey, the incompleteness factor is even larger, $\sim 9$
\citep{Martin2006}.  Extrapolation to all sky, by multiplying the
observed number of dwarfs by these correction factors, is shown by the
thin dashed line in Fig.~\ref{fig:lowmass_projections}.  Our model
falls well below this corrected mass function.

\citet{koposov_etal08} calculate the expected luminosity function of
the faint dwarfs more accurately, by estimating the maximum accessible
volume of the survey.  They conclude that the luminosity function
should rise as $dN/dM_V \propto 10^{0.1 M_V}$, which for a fixed
mass-to-light ratio results in $N(>M_*) \propto M_*^{-0.25}$.  This
estimate is also plotted by a straight line in
Fig.~\ref{fig:lowmass_projections}.  It lies below our first naive
estimate but still significantly above the range of the model.  Both
incompleteness corrections predict over 50 satellites above $10^4\
\Msun$, a factor of two larger than in the model.  Plus, more dwarfs
may remain undetected at larger distances than those probed by the
current surveys ($\sim 300$ kpc).

The discrepancy at low mass persists for all variants of our model and
we were unable to find a set of parameters which adequately reproduced
the stellar mass function at $M_* > 5 \times 10^5\ \Msun$ while
predicting appreciable numbers of $M_* < 5 \times 10^5\ \Msun$
objects. This rather striking result is evidence that our models are not
capturing some critical aspects of the formation of very low mass
galaxies.  The most naive solutions of significantly reducing either
the threshold $\Sigma_{\rm th}$ or the disk structure factor $c$ in
eq. (\ref{eq:r_d}), thereby allowing a larger portion of gas to
participate in star formation, are not viable options since these
modifications invariably overpredict the number of $M_* > 5 \times
10^5\ \Msun$ dwarfs.  In other words, lowering the density threshold
produces so many dwarfs that the ``missing satellites'' are no longer
missing and we are once again left with the expectation that we should
see in the sky $\sim 75$ or more luminous dwarf galaxies around the
Milky Way, whereas we only observe $\sim 30$. \cite{Kang_2008} also 
reports a deficit of ultra-faint dwarfs from an independently-developed 
semi-analytical model of galaxy formation applied to a different 
collisionless $N$-body simulation and using the same cosmic reionization
model employed here.

Observationally, a new interesting puzzle for our understanding of dwarf
galaxy formation is presented by \citet[][]{Ryan_Weber2007}.  In that study
they observe the HI emission from Leo T, the only ultrafaint dwarf
with measurable gas content and recent star formation.  Based on the
density and temperature of HI gas and velocity dispersion of stars,
they find that the gas is everywhere globally Jeans-stable, whereas
the observed pockets of blue, 200 Myr-old stars indicate continuous
star formation.  The observed peak of HI column density is a few
$\Mpc2$, close to the star formation threshold.  Leo T may thus
present another example of stochastic star formation in a handful of
isolated molecular clouds, surrounded by largely inert atomic gas.
Another relatively massive dwarf, Canes Venatici I, also shows a small
fraction of relatively young ($\sim 2$ Gyr), more metal-rich stars in
addition to the predominantly old ($\sim 12$ Gyr), metal-poor
population \citep{martin_etal08}.  Thus even the ultrafaint dwarfs
may, in the future, reveal complex, extended star formation
histories.

\section{Conclusions and Discussion}
  \label{sec:conclusions}

We have presented phenomenological models for star formation histories
of dwarf galaxies in the Local Group, based on the mass assembly
histories in cosmological simulations and the stochastic density
threshold in Kennicutt-Schmidt law of star formation.  Our main
conclusions are as follows:

$\bullet$ Models with a stochastic star formation threshold are much
more successful than non-stochastic models, such as KGK04, in
reproducing the observed star formation histories of the Local Group
dwarfs.  While the KGK04 model predicted 95\% of luminous dwarfs
without any recent ($t < 1$ Gyr ago) star formation, our fiducial
stochastic model correctly predicts that most dwarfs form a few
percent of their stellar mass at late times, in agreement with the
recent star formation fraction inferred for the Local Group dwarfs
(see Table~\ref{tab:sfdata_lg}, $f_{1G}$ column and
Fig.~\ref{fig:sfplots}).  Stochasticity allows star formation to
proceed in isolated regions at late times if the threshold decreases.

$\bullet$ Despite significant improvements of the stochastic model, a
some discrepancies with the data remain.  (1)~Total stellar masses are
typically too large by a factor of several.  This is a generic problem
of both the fixed-threshold and stochastic models presented here.  
(2)~About 10\% of the dwarfs in both fixed-threshold and stochastic models
have anomalously young stellar populations.  These objects build the
bulk of their stellar mass in the last 10~Gyr, whereas all of the
observed dwarfs contain at least 15\% of stars older than 10~Gyr.
These young stellar populations in the models are not created only by
tidally-induced starbursts, but rather represent late mass assembly of
some of the larger satellites.

$\bullet$ Relaxing several model assumptions does not significantly
alter these predictions.  We have considered the following variants of
the fiducial model: allowing late gas accretion within the virial
radius of the host halo; different slopes of the star formation law;
extended epoch of reionization; and different prescriptions for the
photoevaporation of gas from low-mass halos after reionization.  All
of these variants predict statistically similar observables to the
fiducial model.

$\bullet$ A variant of the fiducial model that rejects dwarfs with no
star formation in the first 1 or 2 Gyr after the Big Bang
significantly improves the $f_{10G}$ stellar fraction.  Even though we
do not yet have an adequate justification for such a cut, this model
predicts the $f_{10G}$-distribution fully consistent with the Local
Group data.  Stellar masses, however, are still overestimated.

$\bullet$ Our fiducial model predicts only a modest population of
dwarfs with $M_* \la 10^5\, \Msun$, such as those recently discovered
by SDSS and the MegaCam survey.  The predicted stellar mass function
would be an underestimate if the observed numbers are extrapolated to
all sky.  However, our mass function is still consistent with the
presently known dwarfs.

\medskip

Our phenomenological model contains several free parameters,
which allows significant freedom in the range of predicted properties of 
the satellite galaxies. As we discuss in \S\ref{sec:kgk}, a combination 
of the parameters, $c$ and $\Sigma_{\rm th0}$, is well constrained by the
observed number of dwarfs. Other parameters of the model, apart from the
stochasticity $\epsilon_*$ which strongly affects late-time star formation,
lead only to small and relatively insignificant varations from the fiducial
model. 

This analytical prescription for star formation is applied to the mass 
assembly histories of the halos in the cosmological $N$-body simulation.  
Since we average over three host halos, our results should not depend
significantly on a particular host-halo merger history. Also, by 
construction all the massive satellite halos that become galaxies in 
our model survive tidal disruption in the host halo, so that their total 
number is predicted robustly even if their mass after tidal stripping may
depend on the particulars of the simulation.

We should emphasize that our model is not a unique interpretation of the
SFH data.  Our inferences here neccesarily depend on the assumptions of 
a Schmidt law of star formation and an exponential profile for the gas 
density distribution inside dwarf halos as well as the assumptions we have
made with the minimum density threshold for star formation, 
$\Sigma_{\rm th}$. With this latter part of the model we have found that 
both stochasticity in this threshold over time (\S\ref{sec:threshold})
and a monotonically decreasing function (\S\ref{sec:mono_sec}) can lead to 
extended SFHs -- the feature missing from the KGK04 model.

We have assumed in our model that the gas clouds
moving on circular orbits should generally remain at the same distance 
from the galaxy center. However, during mergers and tidal interactions 
the angular momentum of the gas can be perturbed, leading to radial infall.
Such inflow of gas may bring the central gas density above the threshold 
in some halos soon after the Big Bang and lead to more early star formation.
Modeling this process would be very interesting, as it may provide a nice 
solution for the $f_{10G}$ problem, but the complexity of such a process is
beyond our simple model and requires a detailed hydrodynamic simulation.

Additionally, we have not considered any gas outflows due to the radiative
and thermal feedback of young stars.  Such processes could reduce the amount 
of available gas supply and the total stellar mass of the simulated dwarfs, 
which may lead to a closer agreement of the predicted and observed stellar 
mass function.  However, given the uncertainty in the evolution of the gas 
density profile, we cannot conclude that such feedback is required to 
reconcile the stellar masses.  In fact, from 
Figure~\ref{fig:lowmass_projections} we see that in very low mass halos,
where feedback is expected to be stronger, our model needs a boost, rather
than a reduction, of star formation.  The actual complex details of the 
condensation of molecular clouds in dwarf halos, which lead to the formation
of stars, may turn out more important than the feedback of the stars after
their formation.

In spite of these uncertainties, the conclusion we would like to draw from 
our investigation is that the complex and extended SFHs of the Local Group 
dwarfs are generally consistent with the expected star formation in cold 
dark matter halos, and that this star formation is generally governed by 
the (low) efficiency of conversion of atomic gas into molecular clouds.

\acknowledgements 

We thank Andrew Cole for useful comments on the LMC star formation
data, and Andrey Kravtsov and Todd Thompson for helpful suggestions.
CO thanks the Ohio State University Center for Cosmology and
AstroParticle Physics for its support. OG is supported by the NSF
grant AST-0708087.

\end{document}